\pdfoutput=1

\documentclass[fleqn,usenatbib,usedcolumn,a4paper]{mnras}
\usepackage{natbib}

   \usepackage[british]{babel}             
   \usepackage{newtxtext}                  
   \usepackage[slantedGreek]{newtxmath}    
  
   \let\la=\lesssim 
  
   \usepackage[T1]{fontenc}               
   \usepackage{graphicx}

\newcommand{\msun}{\,\mathrm{M}_{\sun}}
\newcommand{\mpc}{\,\mathrm{Mpc}\,h^{-1}}

\usepackage{graphicx}	
\usepackage{amsmath}	
\usepackage{amssymb}	

\title[Simulation of a high-redshift barred spiral galaxy]{Zoom-in cosmological hydrodynamical simulation of a star-forming barred, spiral galaxy at redshift $z=2$}
\author[F. Vincenzo, C. Kobayashi \& T. Yuan]{Fiorenzo Vincenzo$^{1,2}$\thanks{email: f.vincenzo@bham.ac.uk}, 
Chiaki Kobayashi$^{2}$\thanks{email: c.kobayashi@herts.ac.uk} \& Tiantian Yuan$^{3,4}$\thanks{email: tiantianyuan@swin.edu.au}
\\
$^{1}$School of Physics and Astronomy, University of Birmingham, Edgbaston,  B15 2TT, UK \\ 
$^{2}$Centre for Astrophysics Research, University of Hertfordshire, College Lane, Hatfield, AL10 9AB, UK \\
$^{3}$Centre for Astrophysics and Supercomputing, Swinburne University of Technology, Hawthorn, Victoria 3122, Australia \\
$^{4}$ARC Centre of Excellence for All Sky Astrophysics in 3 Dimensions (ASTRO 3D), Australia }

\begin{document}

\date{Accepted 2019 July 23. Received 2019 July 23; in original form 2019 March 19}

\pagerange{\pageref{firstpage}--\pageref{lastpage}} \pubyear{2019}

\maketitle

\label{firstpage}


\begin{abstract}
We present gas and stellar kinematics of a high-resolution zoom-in cosmological  chemodynamical simulation, which fortuitously captures the formation and evolution of a star-forming barred spiral galaxy, from  redshift $z\sim3$ to $z\sim2$ at the peak of the cosmic star formation rate.
The galaxy disc grows by accreting gas and substructures from the environment.
The spiral pattern becomes fully organised when the gas settles from a thick (with vertical dispersion $\sigma_{v} >$ 50 km/s) to a thin ($\sigma_{v} \sim 25$ km/s) disc component in less than 1 Gyr. 
Our simulated disc galaxy also has a central X-shaped bar, the seed of which formed by the assembly of dense gas-rich clumps by $z \sim 3$.  
The star formation activity in the galaxy mainly happens in the bulge and in several clumps along the spiral arms at all redshifts, with the clumps increasing in number and size as the simulation approaches $z=2$. 
We find that stellar populations with decreasing age are concentrated towards lower galactic latitudes, being more supported by rotation, and having also lower velocity dispersion; furthermore, the stellar populations on the thin disc are the youngest and have the highest average metallicities.
The pattern of the spiral arms rotates like a solid body with a constant angular velocity as a function of radius, which is much lower than the angular velocity of the stars and gas on the thin disc; 
moreover, the angular velocity of the spiral arms steadily increases as function of time, always keeping its radial profile constant. 
The origin of our spiral arms is also discussed.
\end{abstract}




\section{Introduction} \label{sec:intro}

The formation and evolution of the {\it basic} properties of galaxies can be roughly explained in a cosmological context, with the growth followed by the hierarchical clustering of dark matter (DM) halos and feedback from stars and active galactic nuclei (AGNs). However, the formation and evolution of the detailed internal structural properties of galaxies 
have not been well explained yet \citep{conselice2014,dobbs2014}.
The origin of the Hubble sequence \citep{hubble1926} remains a theoretical challenge \citep{benson2010,cen2014,genel2015,clauwens2018}. 
Spiral galaxies are one of the most popular Hubble morphological types in the local Universe. 
(e.g., \citealt{willett2013}).  The essential components of spiral  galaxies include spiral arms, bars, bulges, thin and thick discs.  

The spatially-resolved properties of stellar populations in 
nearby spiral galaxies are obtained with UV, optical, and infrared observations
\citep{rix1993,thornley1996,davis2017,sanchez2017,Yu2018}, while the physical properties of the gas can be investigated in detail by means of 
high-resolution interferometry (e.g. \citealt{walter2008,schinnerer2013,tacconi2013,koribalski2018}); 
to this end, the first observational cycles of the Atacama Large Millimeter Array (ALMA) are providing data of unprecedented 
quality, probing  different phases of the interstellar medium (ISM) 
by making use of the spatial and velocity distributions of 
different molecular species in the galaxy
(e.g., \citealt{faesi2018,sun2018,wilson2018}).

Observations can now resolve internal structures like spiral arms at high redshift ($z=2-3$), an epoch when the Hubble sequence is speculated to emerge \citep{law2012,conselice2014,yuan2017}. The highest spatial resolution observations at high redshift are  usually achieved in small numbers by the technique of gravitational lensing and adaptive-optics aided integral field spectroscopy (IFS). With high accuracy gravitational lensing models, high-redshift galaxies
can be resolved on $\sim$100 pc scales (e.g., \citealt{sharma2018}).
Thanks to ALMA and, in the future, the Square Kilometer Array (SKA), 
we are reaching the sensitivity and resolution power to probe  star-forming disc galaxies at even higher redshifts of $z\gtrsim 3$ (e.g., \citealt{hodge2018}). 
With $\sim$ kpc resolutions, large IFS surveys such as CALIFA \citep{sanchez2016}, MaNGA \citep{bundy2015} and SAMI \citep{croom2012} can provide local baselines of dynamical mapping of thousands of late-type galaxies. Seeing-limited SINFONI \citep{bonnet2004,foster2009} and KMOS \citep{stott2016} surveys  provide  similar  information on a few kpc scales for large numbers of galaxies at $z\sim 1-2$.
In the near future, NIRSpec/JWST \citep{posselt2004,alvesoliveira2018} will provide sub-kpc scale observations on  rest-frame UV and optical properties  of large samples of  galaxies at $z\gtrsim2$. High-resolution cosmological simulations will therefore need to be ready to predict and explain the evolution of galaxy structures at high redshift.

In the past, several works investigated the formation and evolution of star-forming disc galaxies with cosmological simulation techniques, starting -- for example -- from the very early numerical attempts that did not form realistic discs yet \citep{katz1991,navarro2000,Abadi2003}, to more recent efforts that successfully developed 
disc-dominated systems (e.g.,
\citealt{scannapieco2008,agertz2009,agertz2011,guedes2011,aumer2013,stinson2013,hopkins2013,marinacci2014,ubler2014,grand2015,colin2016}). Finally, \citet{grand2017} showed that
cosmological zoom-in simulations can be used to investigate spiral arms and infer their nature at $z\sim0$.

The Hubble classification defines spiral galaxies with or without bars.
The physical processes driving the formation and the growth of such structures, as well as those 
maintaining the stability of the gaseous and the stellar disc of galaxies 
against the pulling forces from both the environment and the feedback from the star formation activity, 
are still a matter of debate  in the astronomical community \citep{dobbs2014,davis2015,schinnerer2017}. 
In summary, the main theories for the origin of spiral arms in disc galaxies are 
the density wave theory \citep{lindblad1960,lin1964,lin1966,kalnajs1971}; the swing amplification mechanism (e.g., \citealt{goldreich1965,julian1966,toomre1981,masset1997,donghia2013}), the multiple mode theory (e.g., \citealt{quillen2011,comparetta2012,sellwoodcarlberg2014}); the 
manifold theory (e.g., \citealt{athanassoula2012,efthymiopoulos2019}), and 
the theory of corotating or dynamical spiral arms (e.g., \citealt{wada2011,grand2012,baba2013}). 
Bars and tidal interactions can also drive spiral density waves on a galatic scale \citep{kormendy1979,salo1993,dobbs2010}, giving rise to the so-called kinematic density waves (see also 
\citealt{kalnajs1973,oh2008,struck2011,oh2015}). Finally, another proposed viable mechanism to develop spiral arms is through the accretion of substructures and gas from the environment into the galaxy gravitational potential  \citep{sellwood1984}. We remark on the fact that almost all simulations have so far found spiral arms as being a transient phenomenon, occurring over a large range of timescales, typically from $\sim1\,\text{Gyr}$ to $\sim10\,\text{Gyr}$ \citep{sellwood2002,fujii2011}.

The different theories for the formation of spiral arms 
predict different characteristic evolution of the dynamical properties 
of the gas and stars on the spiral arms as functions of time. For example, manifold-driven spiral arms have been proven to create an angular velocity pattern which \textit{appears} as constant as a function of time, because -- by assuming a bar co-rotating reference frame -- the trajectories of the particles on the spiral arms are confined in the spiral arms themselves, which line up with the unstable Lagrange points of the bar (the invariant manifolds) \citep{athanassoula2012}; on the other hand, the kinematic density-wave theory -- an other largely favourite theory, giving also rise to a constant angular speed of the spiral arm perturbation as a function of radius -- predict the density waves to show up in the Fourier power spectrum as a power along the inner Lindblad resonance \citep{kalnajs1973}. 
Nevertheless, it is not straightforward to compare these theoretical predictions to observations. In the Milky Way, it is possible to measure the ages of individual stars, from which it is possible to derive the dynamical evolution (e.g., \citealt{rix2013,Bland-Hawthorn2016}), but for external galaxies, observations provide an \textit{instantaneous} snapshot of the properties of the stellar populations and gas 
in the galaxy disc. 

Even if we wanted to compare the spatial distribution and the kinematics of stars with different ages lying in the observed galaxy disc, 
 to determine the best scenario for the formation of the spiral 
 arms, the observed integrated spectrum from a given galaxy region is contributed by a 
mixture of stars with different ages and metallicities. This can be disentangled only 
by making use of stellar population synthesis models, which -- in turn -- strongly 
depend on the assumed initial mass function (IMF), star formation 
history (SFH), stellar evolutionary tracks, and library of stellar spectra. Therefore, the final observational results may strongly depend on the assumptions of models. For this reason, direct comparisons of observed stellar populations with those predicted from spiral arm formation theories are only available very recently with IFS data of nearby galaxies (e.g., \citealt{peterken2018}).

The significant decrease in the observed volume density of spiral galaxies at high redshift implies a close connection between the formation of spiral arms and thin discs \citep{yuan2017}. Most spiral arms in the local Universe reside in a rotating thin disc (vertical height of 200-300 pc) of high angular momentum \citep{epinat2008,glazebrook2013}. 
Studies of  the  Milky Way show that the thin disc of our Galaxy formed around $z\sim 0.8-1$
\citep{freeman2002,haywood2016}. However, whether other spiral galaxies follow the same formation history as the Milky Way is unknown. The question of whether the thin disc formed before or after the thick disc is also contentious \citep{freeman2002,rix2013}. 
Forming a large rotating disc at $z>1$ is theoretically difficult because it takes considerable time to accumulate angular momentum from the accreted halo gas \citep{catelan1996,lago2017}.  Due to limited resolutions, the formation and evolution of large thin discs  with cosmic time  and and their relation to spiral arms are thus far rarely explored in cosmological simulations.

In this paper, we present the first high-resolution chemodynamical zoom-in simulation for the formation of a star-forming barred spiral galaxy at 
high redshift ($z \ge 2$), within a full cosmological framework. 
This allows us to study a disc galaxy that forms and evolves though a large-scale gas accretion, as well as undergoes star formation, feedback, and chemical enrichment within the galaxy.
We characterise in detail the evolution of the physical and kinematical properties of the gas and stellar populations during the formation of the bar and spiral structures in the galaxy.

This paper is structured as follows. In Section \ref{sec:model}, we introduce the basic assumptions of our simulation code, presenting 
both the parent large-volume cosmological simulation and the zoom-in simulation. In Section \ref{sec:results}, we present the results of 
our paper. Finally, in Section \ref{sec:conclusions}, we draw our conclusions.


\section{Simulation model and methods} \label{sec:model}

\subsection{The simulation code} 

We make use of our chemodynamical code \citep{kobayashi2007,vincenzo2018b}, based on \textsc{Gadget-3} \citep{springel2005}, which adopts the smoothed particle 
hydrodynamics (SPH) method to solve the equations of motion of the fluid elements, together with their thermodynamical properties 
\citep{monaghan1992}. 

Our model takes into account three distinct kinds of particles; 
we have \textit{(i)} gas particles, \textit{(ii)} star particles, and \textit{(iii)} dark matter (DM) particles. All these three kinds of 
particles interact with each other 
via their mutual gravitational interaction, and only the physical attributes of gas particles are computed by means of the SPH solver. 
For instance, the main physical attributes of gas particles are their position, velocity, mass, density, temperature, pressure, electron density, 
smoothing length, SFR, and chemical abundances, while the main physical attributes of star particles are their position, velocity, mass, initial mass, formation time, and 
chemical abundances. 

Our scientific target is a young disc galaxy; we do not consider any black hole physics and associated feedback from AGNs.
We address the readers to the work of \citet[and subsequent papers of the same authors]{taylor2014} 
for more details about how black hole physics and AGN feedback have been included in our simulation code, to reproduce the 
observational properties of early-type galaxies and AGN-host galaxies as functions of redshift.

\begin{figure}
\centering
\includegraphics[width=8.0cm]{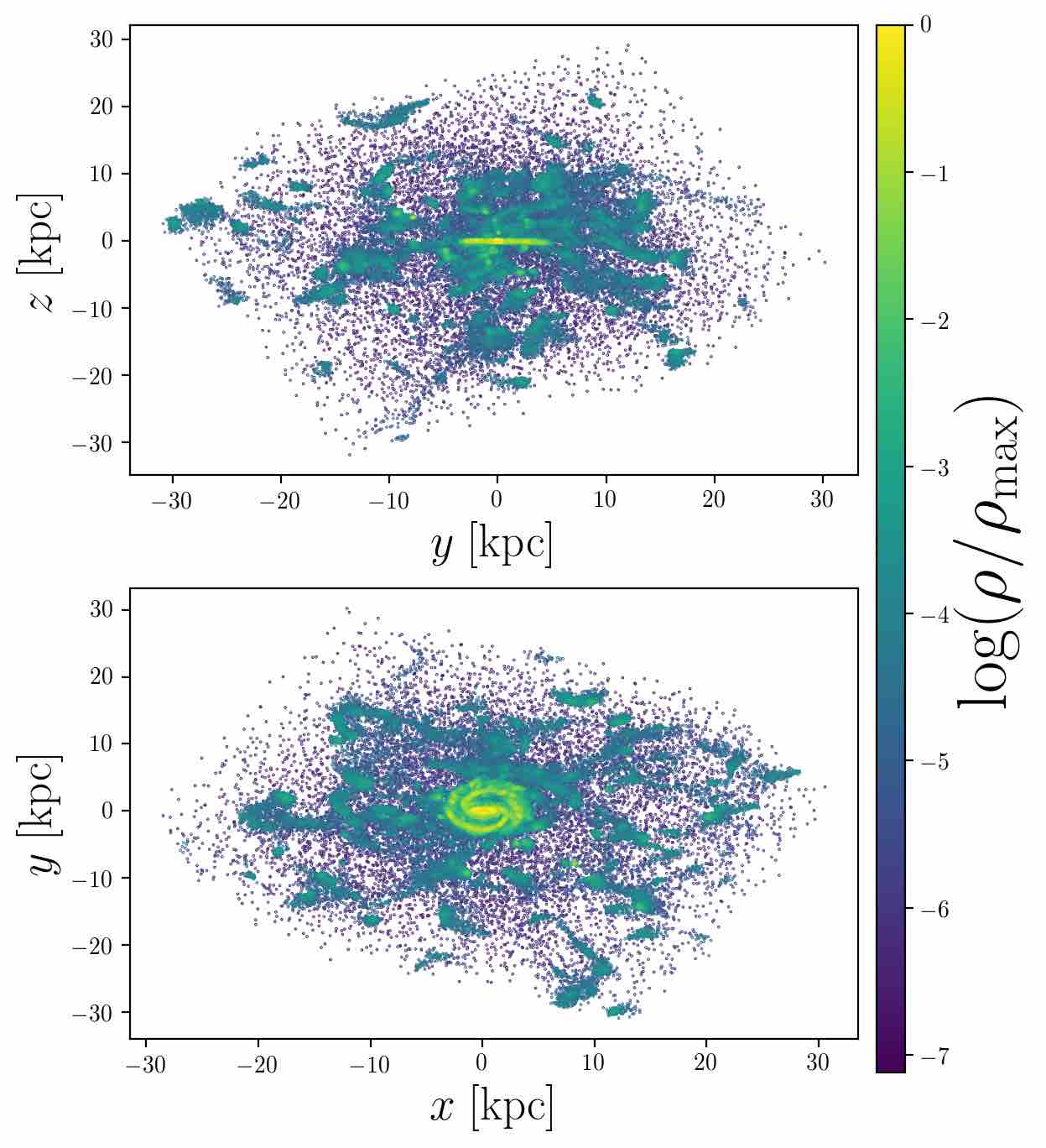} 
\caption{In this figure, we show our simulated star-forming disc galaxy within its closest surrounding environment at redshift $z=2$. In 
the top panel, the galaxy is rotated to be viewed edge-on, while in the bottom panel it is face-on. The colour coding corresponds to the gas density, in 
logarithmic units, normalised to the maximum gas density within the considered region.  }
\label{fig:environment}
\end{figure}

\begin{figure}
\centering
\includegraphics[width=8.0cm]{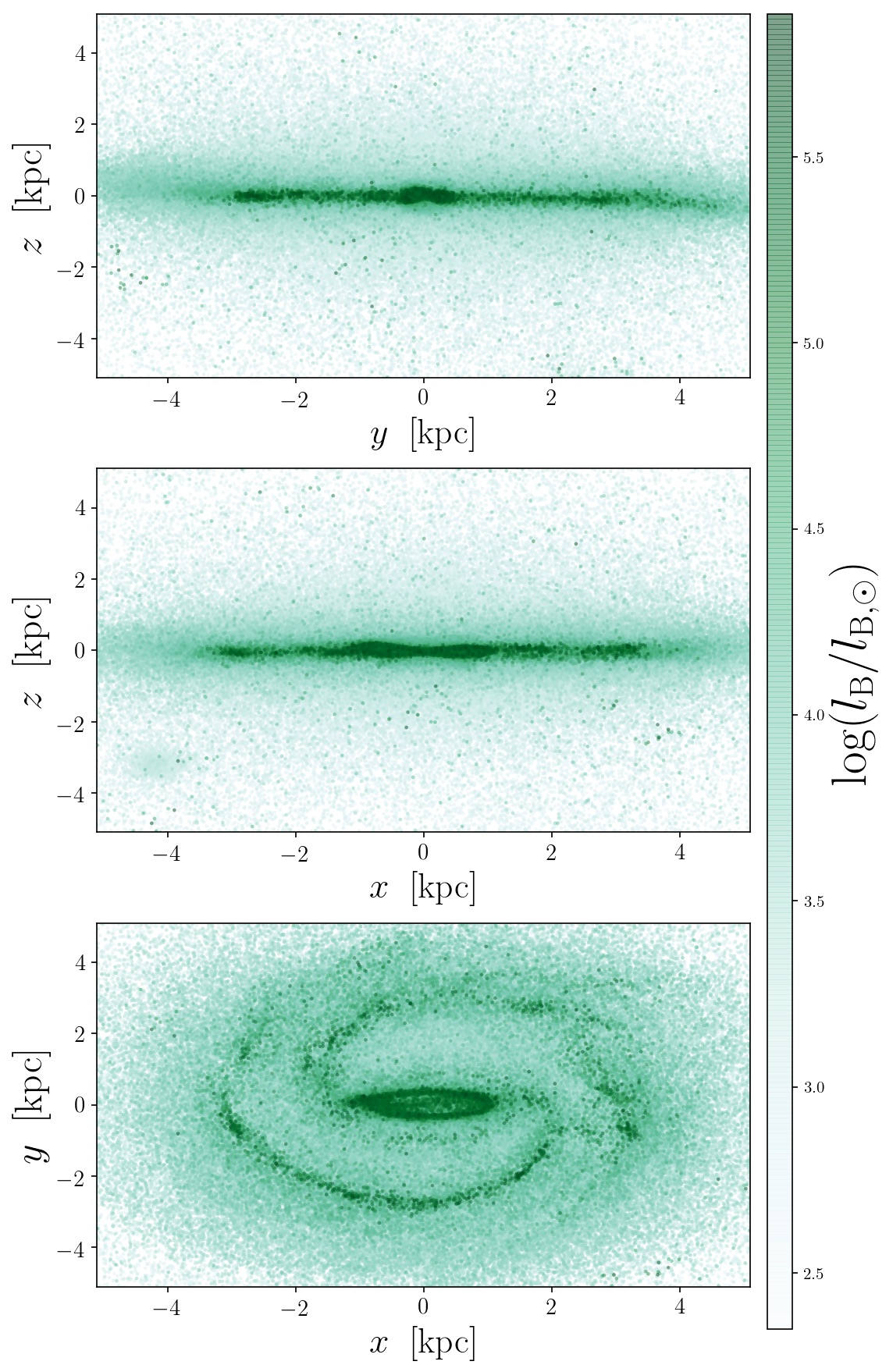} 
\caption{In this figure, we show how our simulated star-forming barred disc galaxy looks like at redshift $z=2$ when its stellar populations are lighted up in the B-band, by using the 
photometric population synthesis model of \citet{vincenzo2016}; in particular, this figure 
contains $845,083$ star particles, which are drawn in the order of their B-band luminosity, $l_{\text{B}}$ , which represents also the colour coding (in logarithmic units, 
normalised to the solar B-band luminosity). }
\label{fig:B-band}
\end{figure}

\begin{figure}
\centering
\includegraphics[width=8.0cm]{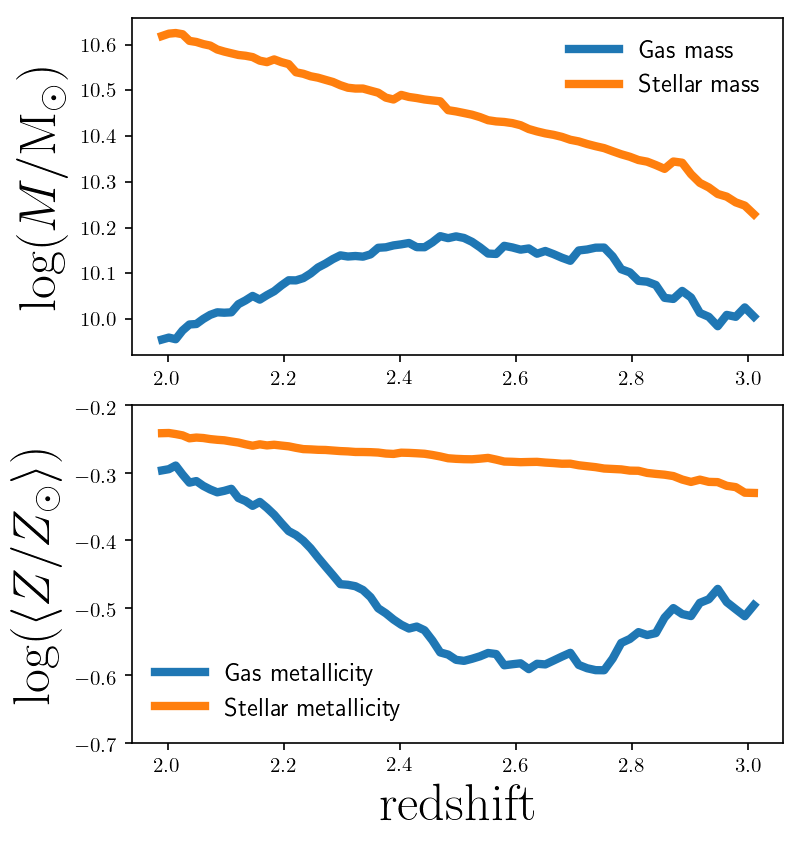} 
\caption{The top panel shows the evolution of the total galaxy stellar and gas mass as functions of redshift. The bottom panel 
shows the redshift evolution of the average stellar and gas-phase metallicities, in logarithmic units, normalised with respect to the 
solar metallicity, $Z_{\sun} = 0.0134$ by \citet{asplund2009}.  }
\label{fig:gal_evolution_tot}
\end{figure}

\begin{figure}
\centering
\includegraphics[width=8.0cm]{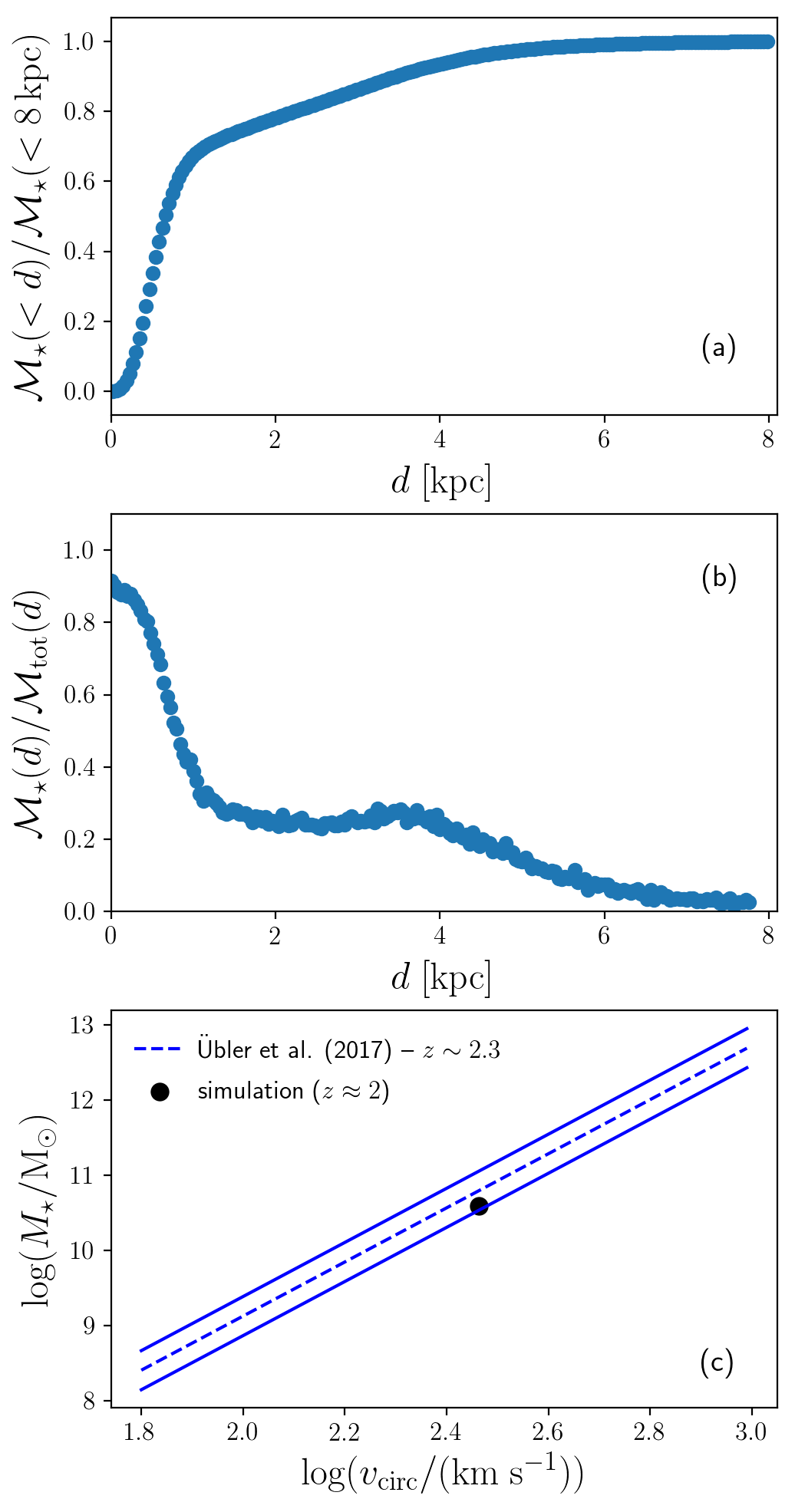} 
\caption{\textit{(a)} Cumulative galaxy stellar mass as a function of the 
galactocentric distance. \textit{(b)} Radial profile of the stellar-to-total mass ratio, where $\mathcal{M}_{\text{tot}}$ includes the DM, star, and gas components. 
\textit{(c)} Observed average Tully-Fisher relation  at redshift 
$z\sim 2.3$ from \citet{ubler2017} is shown as the blue dashed line with the $\pm1\sigma$ deviations as blue solid lines, as compared with the predicted values at $z\approx 2$ in our simulated disc galaxy.  }
\label{fig:mass_profile}
\end{figure}


\subsubsection{Chemical enrichment} 

Our simulation code includes the most detailed chemical enrichment routine, compared with other hydrodynamical codes.
All major stellar nucleosynthetic sources are included, namely 
core-collapse (Type II and hypernovae, HNe) and Type Ia supernovae (SNe), asymptotic giant branch stars (AGBs), and stellar winds from stars of all masses and metallicities.  
We remark on the fact that the chemical abundances associated to a given star particle, $\mathcal{S}_{i}$, correspond to those 
of the gas particle, $\mathcal{G}_{j}$, at the time $\mathcal{S}_{i}$ originated in the past. 

The feedback from the star formation activity 
depends on both the metallicity and age of the star particles.
For the stellar yields and thermal energy feedback, 
we follow the same prescriptions as in \citet{kobayashi2007}, but updated to include failed SNe \citep{vincenzo2018a,vincenzo2018b}. 
For Type Ia SNe, we assume the single-degenerate scenario with metallicity-dependent white dwarfs winds  \citep{kobayashi2009}. 
Moreover, we assume that the nucleosynthetic products and the thermal energy feedback from the ageing star particles in the simulation box are 
distributed to $N_{\mathrm{FB}}=288$ of neighbour gas particles, weighted by the smoothing kernel. 
Finally, the mass spectrum 
of the stars within each star particle defines the so-called initial mass function (IMF), which -- in our model -- 
follows the distribution of \citet{kroupa2008}, very similar to \citet{chabrier2003}.


\subsection{The zoom-in simulation}

\begin{figure*}
\centering
\includegraphics[width=16.0cm]{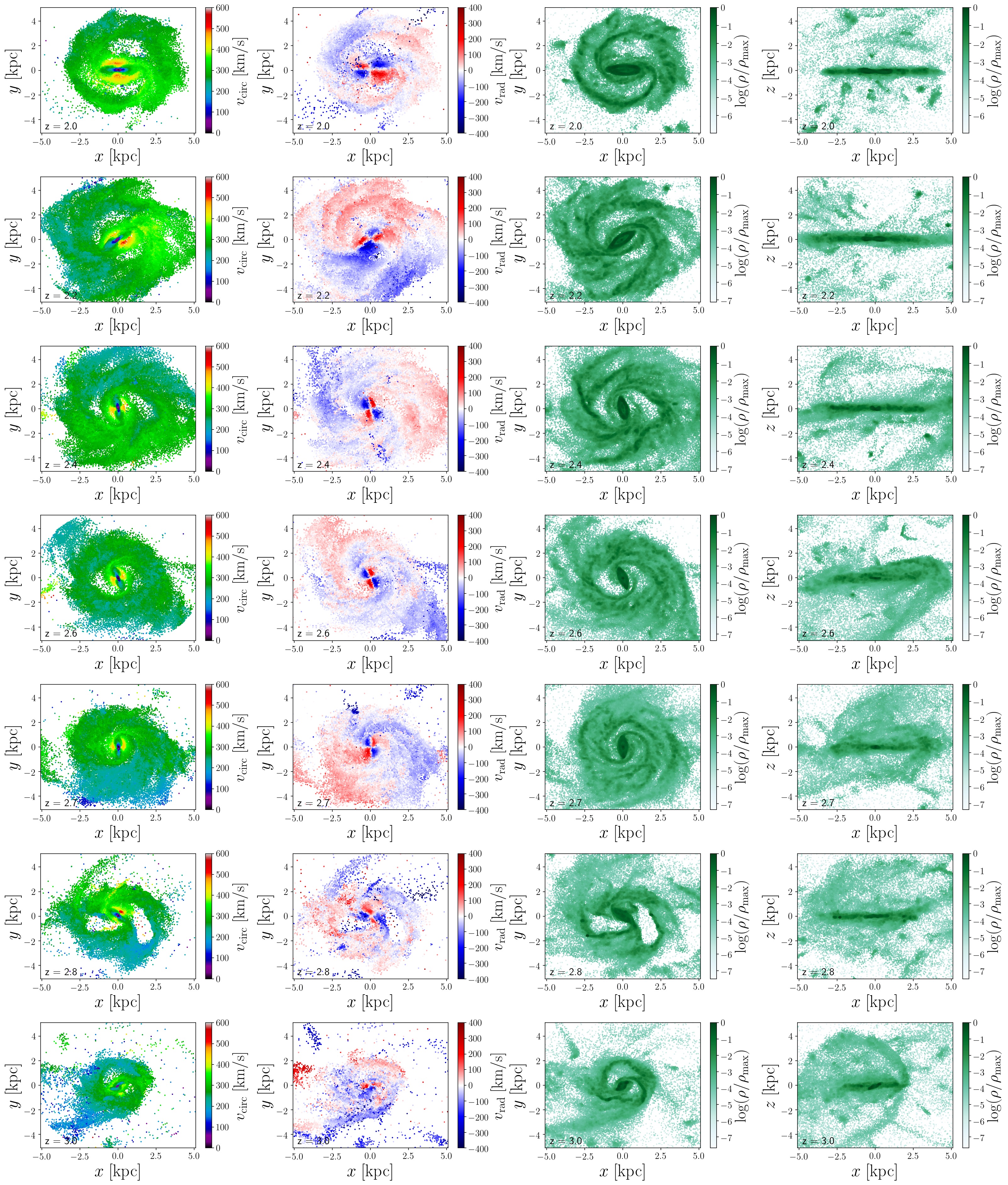} 
\caption{The velocity and density fields within our simulated galaxy as functions of redshift, from $z=3$ to $z=2$ (from the bottom to the top panels).  In particular, from 
left to right, the various columns show similar maps of the simulated galaxy at a given redshift, 
where the colour coding is given by \textit{(i)} the circular velocity of the gas particles (where galaxy is seen face-on), \textit{(ii)} the radial velocity 
of the gas particles (where galaxy is seen face-on), \textit{(iii)} the gas density (where galaxy is seen face-on), and \textit{(iv)} the gas density (where galaxy is seen edge-on).    }
\label{fig:evolution_fields}
\end{figure*}

\begin{figure*}
\centering
\includegraphics[width=16.0cm]{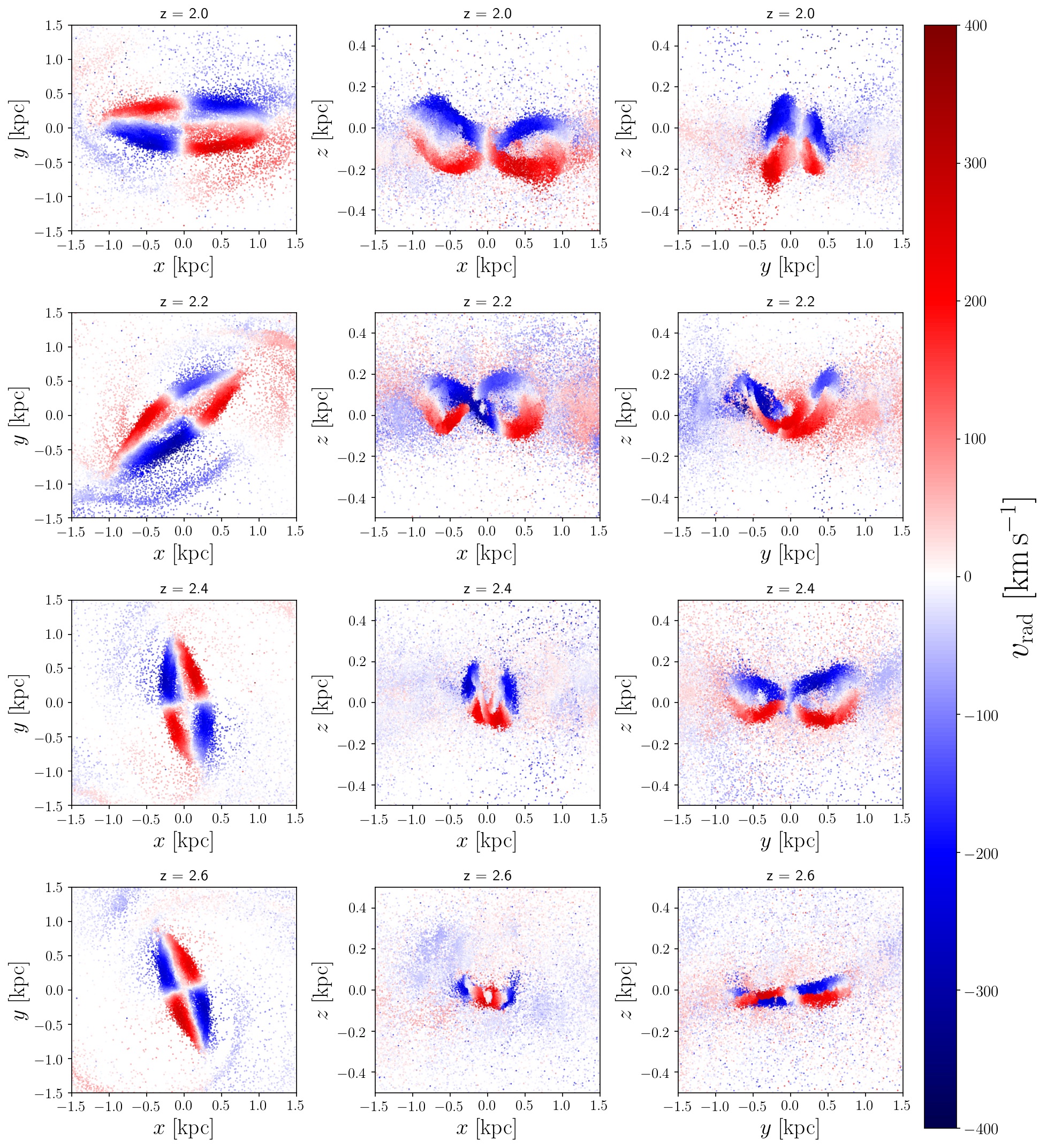} 
\caption{The galaxy bar seen face-on in the $x$-$y$ plane (first column), edge-on in the $x$-$z$ plane (second column), and 
edge-on in the $y$-$z$ plane (third column). From bottom to top, each row of panels shows the galactic bar at a given redshift, with 
the colour coding representing the radial component of the velocity field of the gas particles.   }
\label{fig:bar_vrad}
\end{figure*}

\begin{figure}
\centering
\includegraphics[width=8.0cm]{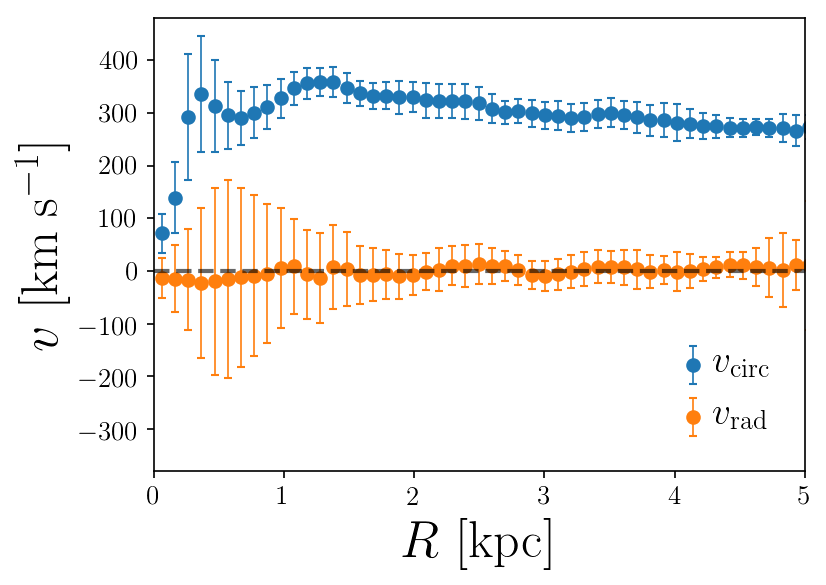} 
\caption{The radial profiles of the circular (blue points) and radial (orange points) velocity patterns of the gas particles in our simulated galaxy as functions of the 
galactocentric distance at $z=2$. }
\label{fig:velocityz2}
\end{figure}

\begin{figure}
\centering
\includegraphics[width=8.0cm]{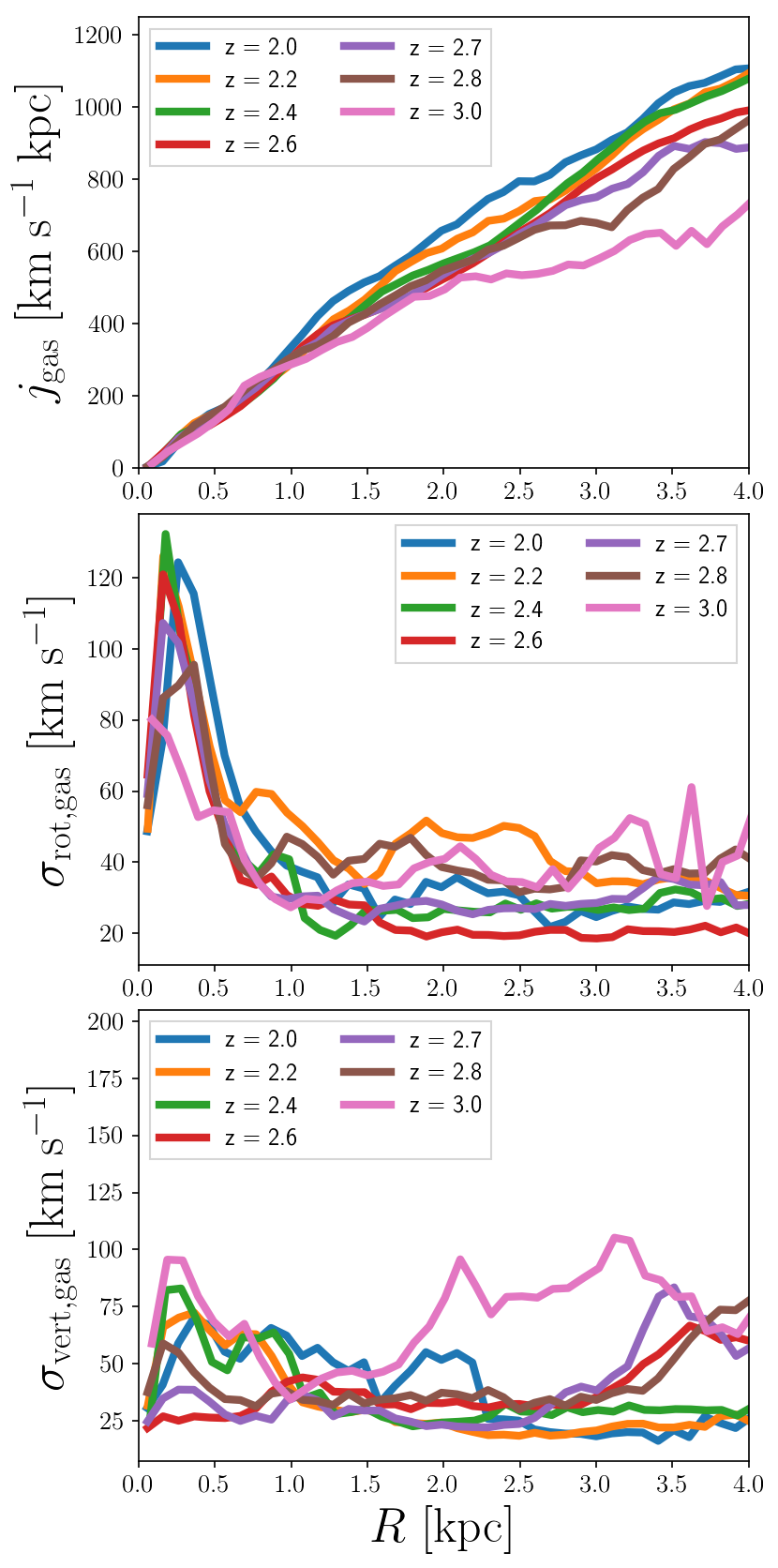} 
\caption{From top to bottom, the various panels show the redshift evolution of the radial profiles of the specific angular momentum, rotational velocity dispersion, 
and vertical velocity dispersion of the gas particles in our simulated galaxy.  }
\label{fig:evolution_j-sigma_gas}
\end{figure}

\begin{figure}
\centering
\includegraphics[width=8.0cm]{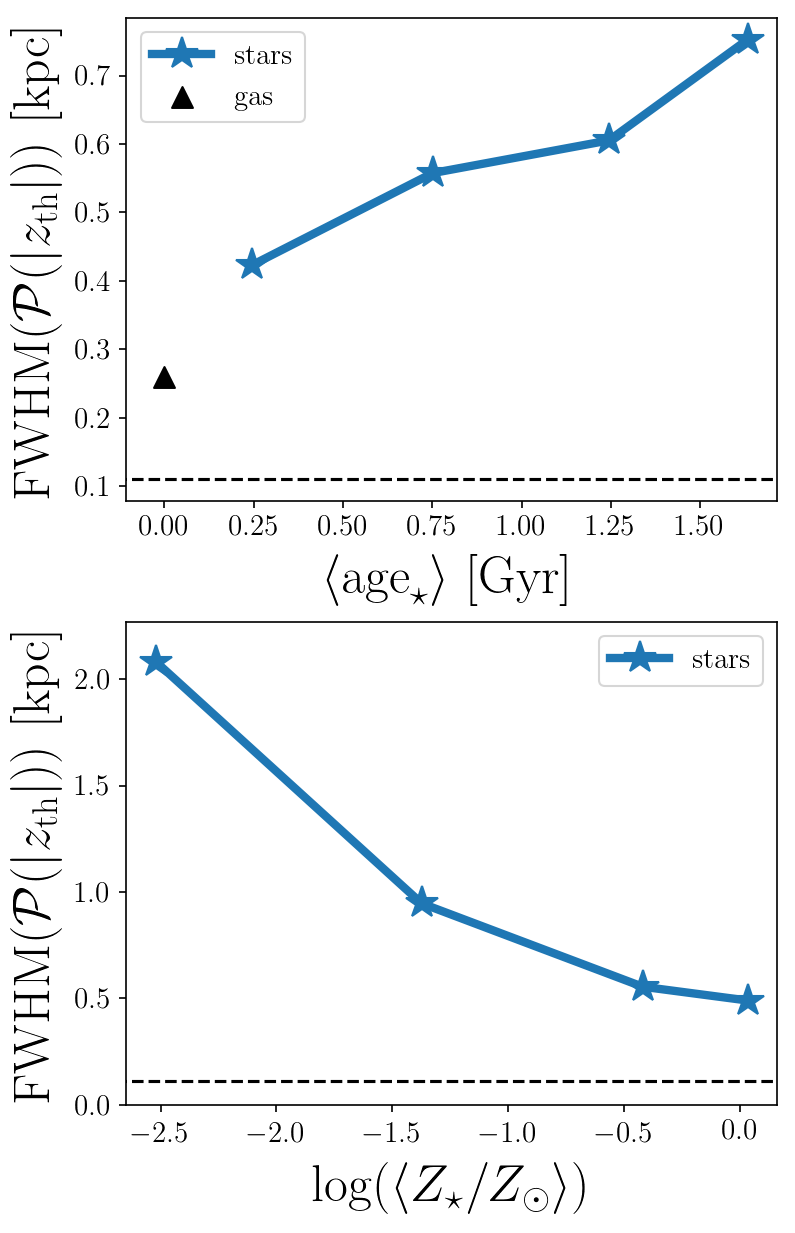} 
\caption{In the top panel, we show how the full width at half maximum (FWHM) of the height distribution 
of the galaxy stellar populations, $\mathcal{P}(\lvert z_{\text{th}} \rvert)$, varies 
as a function of the average stellar age at $z=2$; the black triangle corresponds to the FWHM of the height distribution of the galaxy gas particles, which are 
predicted to lie on a thin disc at $z=2$. The bottom panel shows the FWHM of the height distribution of the galaxy stellar populations as a function of the average 
stellar metallicity. The dashed horizontal line in both panels represents the softening length of our simulation at redshift $z=2$ in physical units ($\epsilon_{\mathrm{gas}} \approx 0.11\;\mathrm{kpc}$). }
\label{fig:thickness}
\end{figure}

\begin{figure}
\centering
\includegraphics[width=8.0cm]{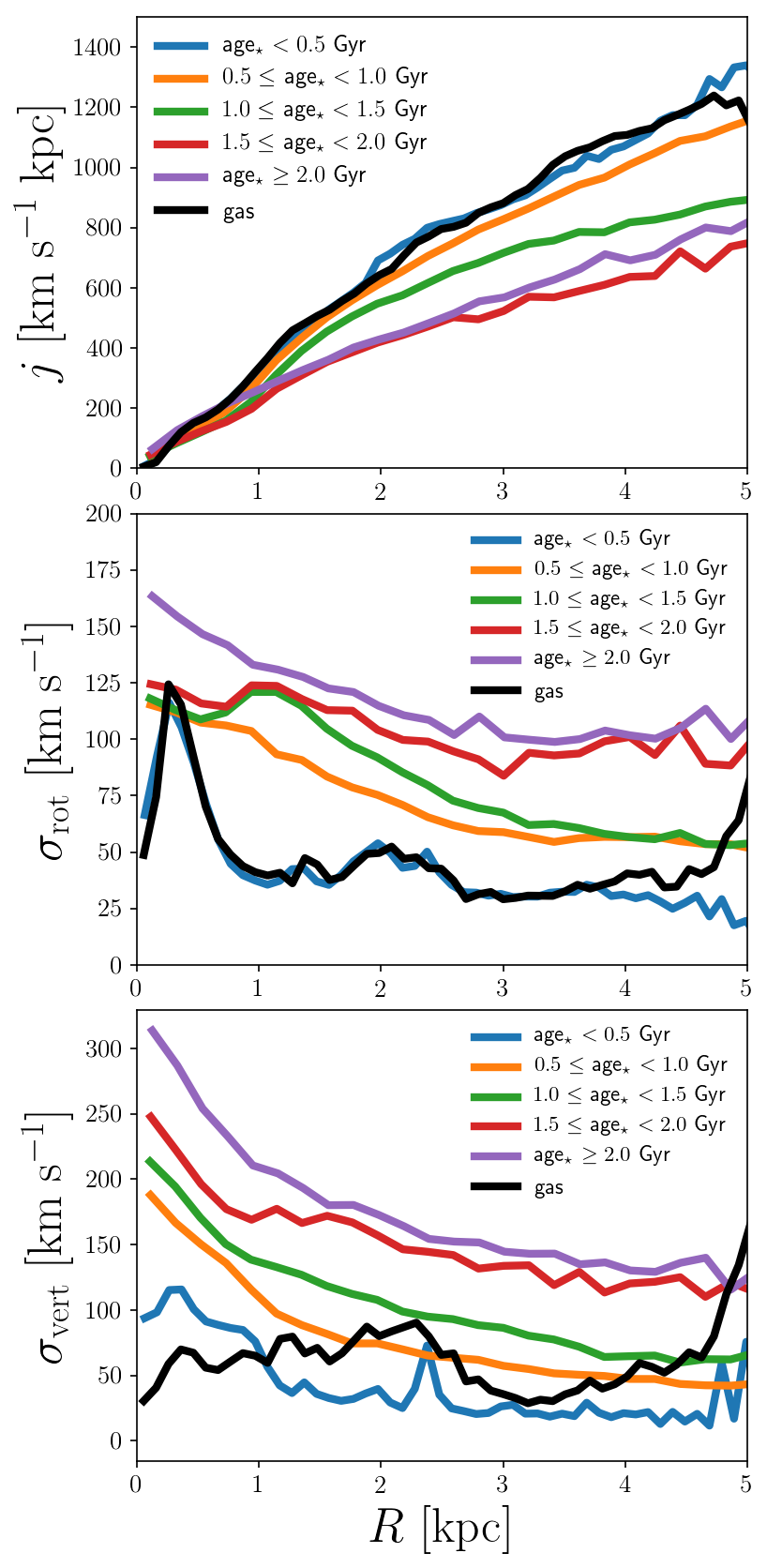} 
\caption{From top to bottom, each panel corresponds to the predicted radial profiles at redshift $z=2$ of the following quantities: 
(\textit{i}) specific angular momentum, (\textit{ii}) rotational velocity dispersion, and 
(\textit{iii}) vertical velocity dispersion. The various curves in each panel correspond to stellar populations in different age bins, with the black curve corresponding 
to the gas particles. }
\label{fig:j-sigma_figure}
\end{figure}

\begin{figure}
\centering
\includegraphics[width=5.0cm]{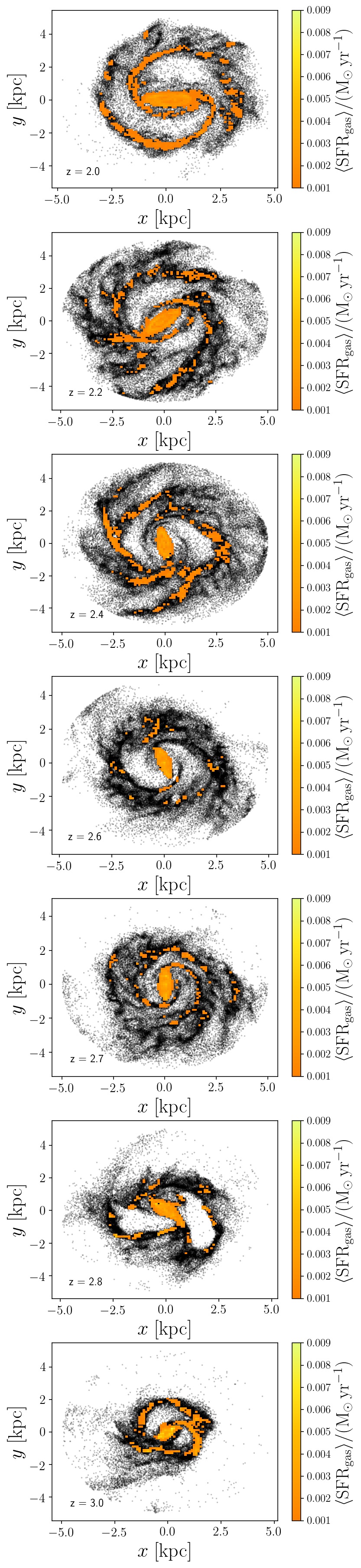} 
\caption{ From bottom to top, the redshift evolution from $z=3$ down to $z=2$ of the SFR map within our simulated galaxy, viewed face-on. 
The black points in the background represent all the gas particles on the disc of our simulated galaxy, while the yellow points highlight the 
gas particles which are star-forming at the given redshift.   }
\label{fig:SFR-maps}
\end{figure}

\begin{figure}
\centering
\includegraphics[width=8.0cm]{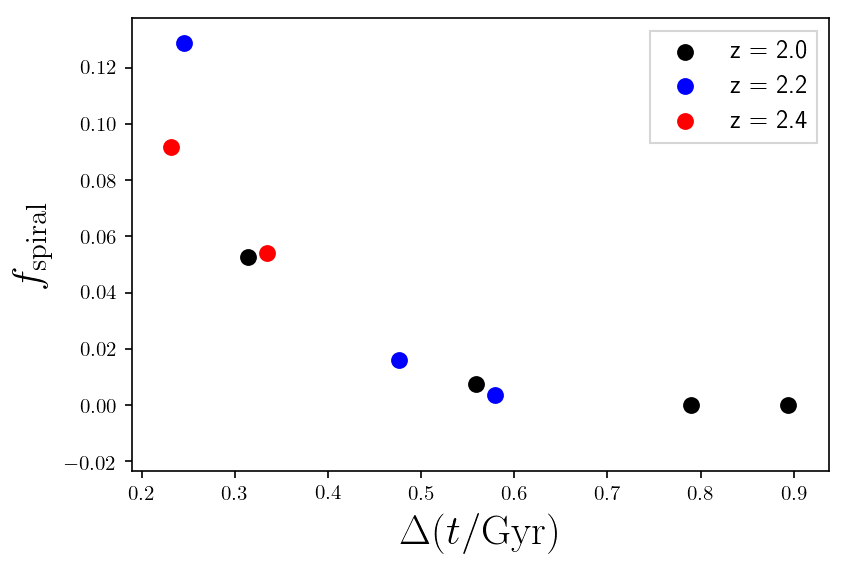} 
\caption{The fraction of star particles, $f_{\text{spiral}}$, which reside on the spiral arms both at redshift $z$ 
(which represents the colour coding of the figure) and at previous epochs 
of the galaxy evolution $z' = z-\Delta z$. In particular, we compute $f_{\text{spiral}}(z)$ for different values of $z'$ in the past, which are then 
translated into values of look-back time.  For example, if we compare the stellar populations on the spiral arms at redshift $z=2$ with those on the 
spiral arms at a time $0.3\,\text{Gyr}$ before $z=2$, we find that they only have $\sim 5$ per cent of stellar populations on the arms in common. }
\label{fig:f_spiral}
\end{figure}

The starting point of this work is represented by a parent cosmological hydrodynamical simulation of a cubic volume of the Universe, 
with comoving side $10\mpc$, and periodic boundary conditions, from which the zoom-in simulation is later set up. 

\textit{The parent cosmological simulation} --- The initial conditions of the parent simulation are drawn by using the \textsc{MUSIC} code\footnote{\url{https://www-n.oca.eu/ohahn/MUSIC/}} 
\citep{hahn2011}. We assume the standard $\Lambda$-cold DM cosmological model, with the following parameters: 
$\Omega_0=0.31$, $\Omega_{\Lambda}=0.69$, $\Omega_{\mathrm{b}}=0.048$, $h = H_{0} / (100 \,\text{km}\,\text{s}^{-1}\,\text{Mpc}^{-1} )  = 0.68$, and 
$\sigma_8 = 0.82$, as given by \citet{planck2016,planck2018}. 
In the parent simulation, we assume a total number of DM and gas particles $N_{\mathrm{DM}} = N_{\mathrm{gas}} = 128^3$, which 
leads to the following mass resolutions in the initial conditions: 
$M_{\mathrm{DM}} \approx3.47\times10^{7}\,h^{-1}\msun$ for the DM particles, and $M_{\mathrm{gas}}=6.35\times10^{6}\,h^{-1}\msun$ 
for the gas particles. The gravitational softnening length is set as $\epsilon_{\mathrm{gas}}\approx0.84\,h^{-1}\;\mathrm{kpc}$ in comoving units.


\textit{The target DM halo} --- 
In summary, a target DM halo is selected in the parent cosmological 
simulation at redshift $z=0$, and then re-simulated from the initial conditions (at redshift $z=49$) with a much larger number of particles, giving rise to 
a zoom-in cosmological simulation. 
Our target DM halo in the parent simulation is selected, because it is fairly isolated, it has few substructures, 
and it lies within a less dense environment than the other DM halos in the simulated cosmological volume. 
By considering the mass within the virial radius, which is $r_{\text{vir}}\approx259\,\text{kpc}/h$, 
the target DM halo in the parent simulation 
has, at redshift $z=0$, a total mass $M_{\text{DM,h}} = 3.35\times10^{12}\,h^{-1}\msun$ in the DM component, total stellar mass 
$M_{\star,\text{h}} = 1.78\times10^{11}\,h^{-1}\msun$, and total gas mass $M_{\text{gas,h}} = 3.76\times10^{11}\,h^{-1}\msun$. By looking at the central galaxy in the target DM halo of our parent cosmological simulation at redshift $z=0$, its SFH in the last $3$ Gyr has been steadily quenched as a function of time.     

\textit{The initial conditions of the zoom-in simulation} ---  The initial conditions of the zoom-in simulation are 
drawn by using the \textsc{MUSIC} code \citep{hahn2011}. 
Firstly, we select at redshift $z=0$ all the DM particles of the target DM halo which lie within a spherical region defined 
by a ``zoom-in radius'', $r_{\text{zi}}$, 
such that $\langle \rho( r_{\text{zi}} ) \rangle \approx 200 \times \rho_{\text{crit}} $, where 
$\rho_{\text{crit}}$ is the critical density of the Universe. 
We have checked that there is no artificial boundary effect, namely 
the DM and gas particles in the less dense regions outside the ``zoom-in sphere'' remained well 
in the outskirts of the target halo also at higher redshifts. 
We then determine the region in the initial conditions of the parent simulation (corresponding to 
redshift $z=49$) spanned by all the DM particles in the zoom-in sphere at $z=0$. Considering only this 
zoom-in region, we make use of the MUSIC code to draw new density and velocity 
fields in the initial conditions, which sample the original fields of the parent simulation but with a larger number of resolution elements. 

\textit{The zoom-in simulation} --- We develop a new zoom-in cosmological chemodynamical simulation, which has vacuum boundary conditions, a total number of DM and gas particles 
$N_{\text{DM}} = N_{\text{gas}} = 5,052,912$, which results in mass resolutions $M_{\text{DM}} = 5.42\times10^{5}\,h^{-1}\msun$ for the DM particles, and 
$M_{\text{gas}} = 9.92\times10^{4}\,h^{-1}\msun$ for the gas particles in the initial conditions. The cosmological parameters are the same as in the 
parent simulation. The gravitational softening length is 
$\epsilon_{\mathrm{gas}}\approx0.231\,h^{-1}\;\mathrm{kpc}$ in comoving units, which -- at redshift $z = 2$ -- corresponds to $\epsilon_{\mathrm{gas}} \approx 0.11\;\mathrm{kpc}$ in physical units. 
We run our zoom-in simulation from $z=49$ down to $z=1.8$ with vacuum boundary conditions.


\section{Results} \label{sec:results}

\subsection{Basic properties of our zoom-in galaxy}

In Fig. \ref{fig:environment}, we show the outcome of our zoom-in simulation at redshift $z=2$, by focusing on the main central galaxy in the simulated volume, 
together with its closest surrounding environment. Each point in the figure corresponds to a gas particle, 
and the colour coding represents the gas density, in logarithmic units, normalised to the maximum gas density in the considered region. 
The region has been rotated so that the central galaxy can be seen edge-on (top panel) and face-on (bottom panel). 
By looking at Fig. \ref{fig:environment}, many substructures are present in the galaxy halo, which represent the gas reservoir from which 
the central galaxy continuously grows in mass as a function of time, fuelling active star formation and chemical enrichment processes.

In the simulated volume, the formation and growth of a central star-forming barred spiral galaxy begins before redshift $z\approx3.0$ continuing down to redshift 
$z \approx 1.8$, corresponding to a time interval of $\approx 1.48\,\text{Gyr}$. The main dynamical features of this galaxy are clear 
both in the gas and in the stellar components. 
For instance, Fig. \ref{fig:B-band} shows the simulated disc galaxy in the B-band luminosity at redshift $z=2$, 
which highlights the young stellar populations in the galaxy. 
In the various panels of Fig. \ref{fig:B-band}, the galaxy is drawn edge-on in the plane $y$-$z$ (top panel), 
edge-on in the plane $x$-$z$ (middle panel), and face-on in the plane $x$-$y$ (bottom panel). 

We quantify the disc structure of our simulated galaxy  by measuring the disc scale length and scale height in the mocked V-band image. We find that the V-band half-light radius of the stellar disc of our simulated galaxy is $r_{h,\text{V}} = 2.97\,\text{kpc}$, and the vertical scale height (by fitting an exponential function to the vertical luminosity profile; \citealt{kregel2005}) is  $z_{h}=0.36\,\text{kpc}$. Note that the vertical height of the disc depends on the age of the stellar populations, with  younger stars and gas residing in a thinner disc (Section 3.3). 
From a first glance at the results of our simulation, we find the presence of two spiral arms, departing from the edges of a central bar, which has 
a major axis $a_{\text{bar}} \approx 1.4\,\text{kpc}$ and minor axis $b_{\text{bar}} \approx 0.3\,\text{kpc}$. 
Since the two spiral arms point, at the outskirts, towards an opposite direction with the respect to the motion due to the 
galaxy rotation, we conclude that the simulated disc galaxy has trailing spiral arms.  
Hence, our zoom-in galaxy is a barred spiral galaxy, although we did not choose the initial conditions or did not tune any parameters of baryon physics in our simulation code.

At high redshifts, forming a stable and persistent disc represents a challenge for galaxy formation and evolution models 
embedded within a cosmological framework, because of the more turbulent physical conditions of the environment than in the local Universe. 
Moreover, at high redshifts, galaxies typically cover smaller physical spatial distances and have lower masses than nearby galaxies, making the disc structures 
more fragile. Therefore, in order not to enhance abruptly the SFR within the galaxy, which would dramatically heat the disc, 
the accretion of (tidal) substructures and gas from the filaments and the halo needs be very smooth with time, as well as the star formation history needs to be gradual 
and not bursty as function of time. Finally, 
the disc galaxy should reside in low density cosmic regions, to avoid major merger events, or high-velocity encounters, which also may make the disc unstable from a dynamical 
point of view (e.g., \citealt{cen2014}). 

In our zoom-in simulation, we first witness  the formation of a dense clump of gas and stars at redshift $z\approx3.5$, as a consequence of the assembly of gas-rich and compact stellar systems. This clump 
represents the ``seed'' of what will later become 
the central bar of the simulated disc galaxy.  Then, as aforementioned, by redshift $z\approx3$, a rotating galaxy disc begins to grow with time. 
In the top panel of Fig. \ref{fig:gal_evolution_tot}, we show the redshift evolution of the total stellar and gas masses of our simulated disc galaxy, while the bottom panel 
shows the evolution of the stellar and gas-phase metallicities as functions of redshift. At redshift $z=2$, the simulated galaxy has total stellar and gas masses 
$M_{\star} = 4.16 \times 10^{10}\,\text{M}_{\sun}$ and $M_{\text{gas}} = 9.03 \times 10^{9}\,\text{M}_{\sun}$, respectively. The average galaxy stellar and gas 
metallicities at $z=2$ are $\log(Z_{\star}/Z_{\sun}) = -0.24$ and $\log(Z_{\text{gas}}/Z_{\sun}) = -0.30$. To compute the average integrated galaxy properties, we firstly fit 
the distribution of the gas particles along the $x$-, $y$-, and $z$-axis with Gaussian functions, and then we consider only the particles which lie within $4$-$\sigma$ of the 
three fitting Gaussians. 

Both the total galaxy 
stellar mass and the average stellar metallicity in Fig. \ref{fig:gal_evolution_tot} continuously increase as functions of time, 
without any visible sudden increase or decrease, meaning that there are no major merger events in the considered redshift interval. 
Concerning the total galaxy gas mass, in the first stage of the galaxy evolution, from $z=3$ to $z\approx2.5$, 
$M_{\text{gas}}$ smoothly increases as a function of time, because of gas accretion from the intergalactic medium. 
The accreted gas is of primordial chemical composition, and the average gas-phase metallicity in the galaxy, $\langle Z_{\text{gas}} \rangle$, decreases in this stage. This means that the accretion process 
dominates over the star formation and chemical enrichment processes inside the galaxy, whose main effects are 
to consume the gas and deposit metals in the ISM after the star formation activity. 
From $z\approx2.5$ to $z=2$, on the other hand, $M_{\text{gas}}$ smoothly decreases as a function of time, and the average galaxy 
gas-phase metallicity $\langle Z_{\text{gas}} \rangle$ increases, because the star formation process inside the galaxy 
dominates in this stage over the accretion of gas from the intergalactic medium. 

In Fig. \ref{fig:mass_profile}(a-b), we show how the stellar mass is distributed within our simulated disc galaxy. In particular, the radial profile of the cumulative galaxy stellar mass is shown in Fig. \ref{fig:mass_profile}(a), and the stellar-to-total mass ratio as a function of the galactocentric distance is shown in Fig. \ref{fig:mass_profile}(b), where 
the total mass, $\mathcal{M}_{\text{tot}}$, is defined as the sum of the DM, star, and gas galaxy mass components. In Fig. \ref{fig:mass_profile}(c), we compare the observed Tully-Fisher relation with the predictions of our simulated galaxy.  The observational data (blue lines) are from \citet{ubler2017}  for a sample of star-forming disc galaxies at redshift $z\approx 2.3$ in the KMOS$^{3\text{D}}$ survey, and the black point corresponds to our simulated disc galaxy at redshift $z=2$. 

By looking at Fig. \ref{fig:mass_profile}(a), we find that approximately $70$ per cent of the galaxy stellar mass is concentrated in the galaxy bulge, with the remaining $30$ per cent contributing to the galaxy stellar disc; this gives rise to a relatively high bulge-to-disc (B/D) mass ratio, which is about $\sim 2.3$, a value typical of the earliest type spirals at $z \sim 0$ (see, for example, \citealt{graham2008}),  though there are no available observational data for the B/D ratio for a large sample of high-redshift disc galaxies of different morphological type.

 Interestingly, we find that the stellar-to-total mass ratio, $\mathcal{M}_{\star}/\mathcal{M}_{\text{tot}}$, is almost constant on the galaxy disc about a value of $\sim0.25$-$0.30$ (see Fig. \ref{fig:mass_profile}b). This means that, in the annulii at galactocentric distances between $2$ and $4\,\text{kpc}$, stars and gas together almost equally contribute to the total galaxy mass as the DM. Our predicted disc-to-total mass ratio is almost one order of magnitude larger than, for example, the assumed value in the simulation of \citet{hu2016} for an isolated MW-like galaxy. In our simulated galaxy, the dynamical evolution of the spiral structure on the disc may take place in a regime in which self-gravity is important.
 We also note that the baryon fraction in our bulge is as large as in the nearby early-type galaxies \citep{cappellari2016}.

 Finally, in Fig. \ref{fig:mass_profile}(c), we show that our simulated disc galaxy follows the observed Tully-Fisher relation in the same redshift range. Even though the Tully-Fisher relation involves integrated quantities, the qualitative agreement between the observations (B/D mass ratio and Tully-Fisher relation) and our simulation may suggest that our simulated galaxy does not heavily suffer from the overcooling problem \citep[e.g.][]{steinmetz1999}

\subsection{Kinematical properties of the gas on the galaxy disc} \label{sec:gas}

In Fig. \ref{fig:evolution_fields}, from the bottom panel to the top panel, we show how the galaxy velocity and density fields evolve from $z=3$ to $z=2$, after both the disc and the bar are formed. 
The first column shows the rotational velocity field, which is the colour coding in the figure, and the galaxy is viewed face-on; 
the second column the radial velocity fields (the galaxy is face-on), and the third and fourth 
columns show the gas densities within the galaxy, viewed face-on and edge-on, respectively. 
We remark on the fact that, in Fig. \ref{fig:evolution_fields}, from bottom to top, 
the galaxy rotates counterclockwise as a function of time. 

Depending on whether we draw the rotation curve along the major or minor axis of the bar, there are differences in the radial profiles 
of the gas rotational velocity. In fact, by looking at the first column of Fig. \ref{fig:evolution_fields}, 
there is a bump in $v_{\mathrm{circ}}$ next to the bar, along the minor axis; this is due to the bar rotation itself, which increases 
the gas kinetic energy, both downstream and upstream with respect to the bar rotation. 
Moreover, we find that the bar has an X-shaped structure, and that the gas particles in the bar are on figure-of-eight orbits \citep{binney2008}; this can be better appreciated 
by looking at Fig. \ref{fig:bar_vrad}, where only the bar region is zoomed 
at different redshifts (from top to bottom, one moves towards higher redshifts), where the bar is viewed face-on in the first column, 
and edge-on in the third and fourth columns. The colour coding in all panels of Fig. \ref{fig:bar_vrad} represents 
the radial component of the gas velocity field. 

In Fig. \ref{fig:velocityz2}, the radial profile of the circular velocity of the gas at redshift $z=2$ (in blue) is compared 
with the radial profile of the radial velocity of the gas (in orange). 
By looking at the circular velocity profile, there is a linear increase in the innermost ($\sim$0.5 kpc) galaxy regions, corresponding to the location of the bar; 
this is a clear signature of the fact that the bar rotates like a solid body. 
The linear increase of the circular velocity profile is then followed by a flattening around a mean value 
$\langle v_{\text{circ}} \rangle = 287.0 \pm 12.3 \,\text{km/s}$, which is 
computed between $R = 2.5$ and $5\,\text{kpc}$ from the galaxy centre. 
Note that this rotational velocity is faster than in our Milky Way, which is $\sim 4$ times more massive and much more evolved than our simulated galaxy. This is due to the fact that the MW experienced  mild gas accretion and  star formation activity in the last $\sim 5$ Gyr, building up most of its stellar mass over an extended period of time \citep{kobayashi2011a}. 

Finally, there is a large dispersion of the radial velocities of the gas 
in the bar region (see also Fig. \ref{fig:bar_vrad}), but both $v_{\text{rad}}$ and the 
dispersion of $v_{\text{rad}}$ become 
low on the disc, where $\langle v_{\text{rad}} \rangle = 3.9 \pm 5.8 \,\text{km/s}$, 
which is computed again between $R=2.5$ and $5\,\text{kpc}$ from the galaxy centre.

In Fig. \ref{fig:evolution_j-sigma_gas}, we show how the main kinematical properties of the gas particles on the galaxy disc are predicted to 
evolve as functions of redshift; in particular, we show the evolution of the radial profile of the specific angular momentum of the gas (top panel), the evolution of the 
gas rotational velocity dispersion (middle panel), and the evolution of the gas vertical velocity 
dispersion (bottom panel). To make Fig. \ref{fig:evolution_j-sigma_gas}, we consider 
only the gas particles with heights $\lvert z_{\text{th}} \rvert < 0.4\,\text{kpc}$, above or below the disc. 

Galaxies tend to minimise their energy by concentrating their mass towards the centre, and redistributing 
angular momentum and hence kinetic energy outwards.  
By looking at Fig. \ref{fig:evolution_j-sigma_gas}, the specific angular 
momentum, $j_{\text{gas}}$, monotonically increases as a function of the galactocentric distance on the disc; 
this is due to the flat rotation curve in the outer disc, which is predicted at almost all 
redshifts in our simulated galaxy. 
Interestingly, we predict $j_{\text{gas}}$ to increase, on average, as a function of redshift, for any given galactocentric distance on the disc. 
This constant increase of $j_{\text{gas}}$ with redshift is due to the kinetic energy 
deposited onto the disc by the gas and substructures accreted from the intergalactic medium, which make the disc growing in mass and 
size as a function of time. 
In particular, the half stellar-mass radii of our simulated galaxy at $z=2$, $2.4$, $2.8$ and $3$ are $r_{h,M} = 2.19$, $1.99$, $1.53$, and $1.41$ kpc, respectively. 

\begin{figure}
\centering
\includegraphics[width=8.0cm]{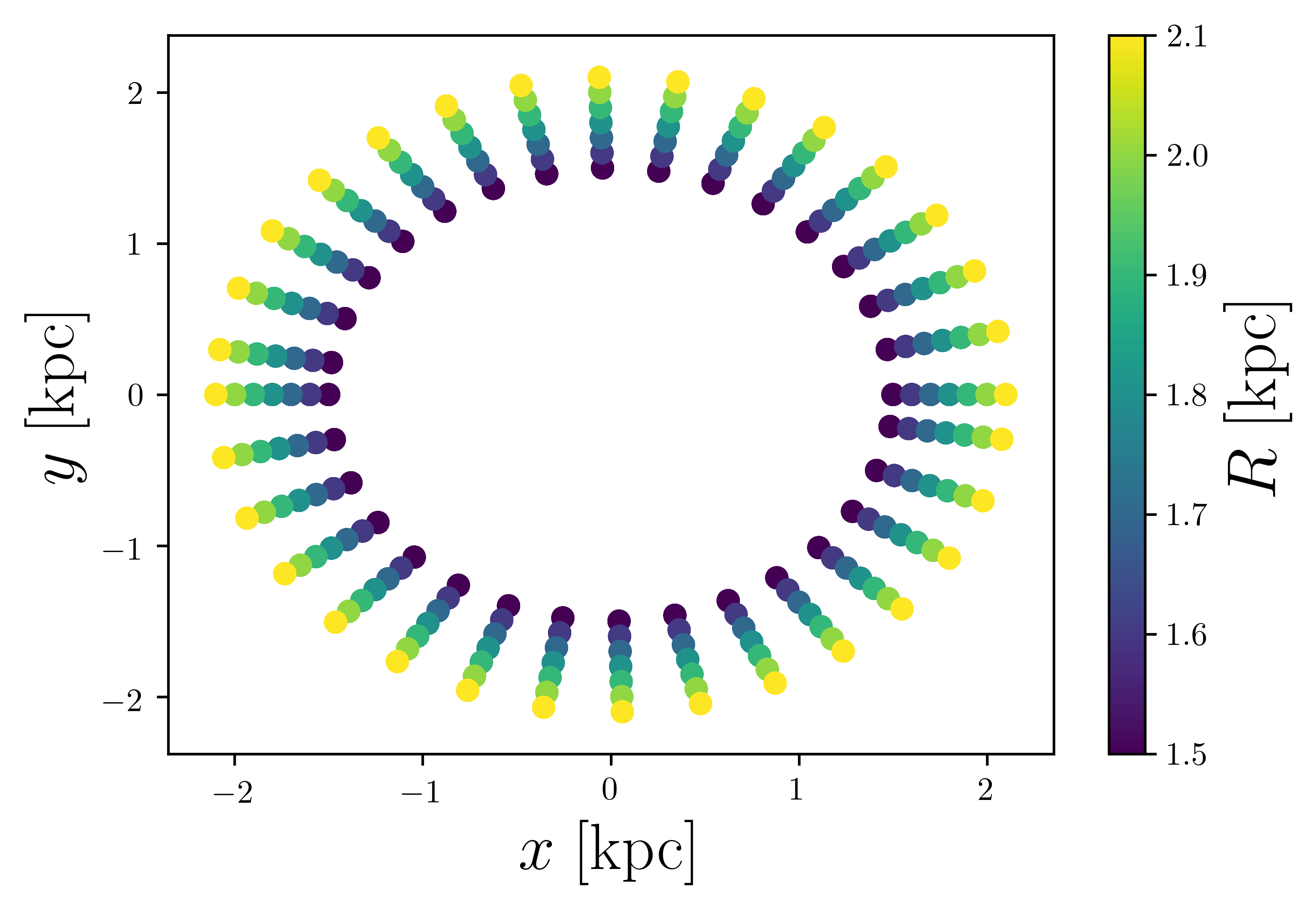} 
\caption{In this figure, we show the position of the ideal observers, sitting at rest on the galaxy disc, that we use to register the passage of the density perturbation driven by the spiral arms, by considering all available snapshots between redshift $z\approx1.8$ to $z\approx2.6$. Each individual observer defines a box with size of $0.1\,\text{kpc}$ around them.  }
\label{fig:eobservers}
\end{figure}

The radial profiles of both the rotational and vertical velocity dispersions (middle and bottom panels of Fig. \ref{fig:evolution_j-sigma_gas}, respectively) 
are predicted be $\lesssim 60\,\text{km/s}$ on the disc. 
The central peak in $\sigma_{\text{rot}}$ is due to the bar, which -- as aforementioned -- 
has a strong effect on the gas kinematics, which is seen in our simulation in terms of a significant increase of 
the dispersion of the rotational and radial velocity components. 
The vertical velocity dispersion is not affected by the bar, 
being more sensitive to the physical conditions of the environment. 
Since the disc at $z=3$ has a smaller physical size than at $z=2$, the high values 
of $\sigma_{\text{vert}}$ at large $R$ at $z=3$ correspond to regions outside the main galaxy body. Fig. \ref{fig:evolution_j-sigma_gas} also demonstrates that in our spiral galaxy, the thick disc formed before the thin disc. 

\begin{figure*}
\centering
\includegraphics[width=12.0cm]{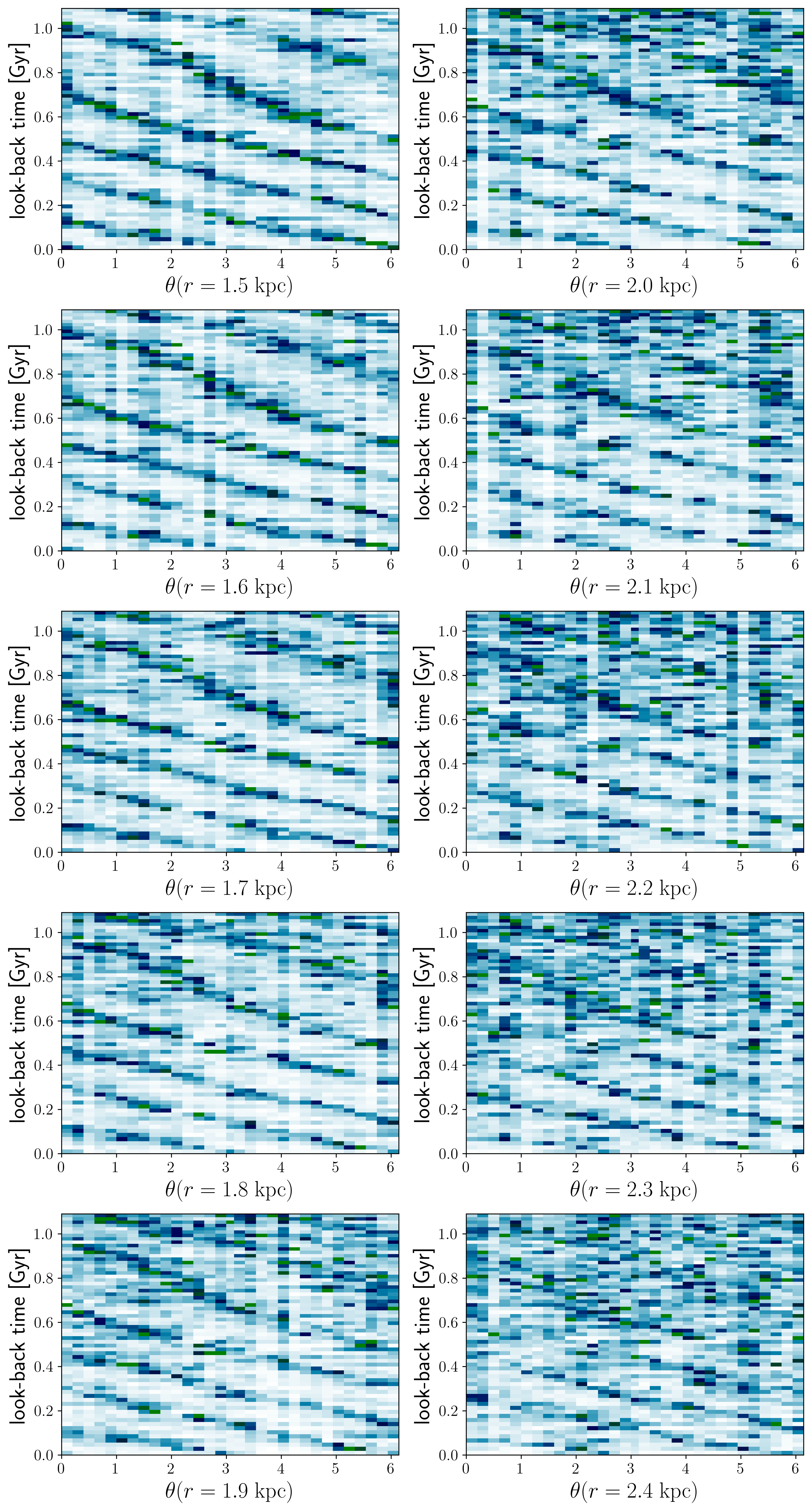} 
\caption{The time evolution of the train of impulses in the gas density, due to the passage of the spiral arms, as seen by each observer of Fig. \ref{fig:eobservers}, according to its azimuth angle $\theta$. Each panel corresponds to the galactocentric distance of the observers, and the colour coding goes from $0$ to $1$, being normalised to the maximum number of particles (the maximum amplitude of the impulses) that each observer has ever seen passing through their region. We note that this figure is made by considering all snapshots available from redshift 
$z\approx 1.8$ (look-back time $t_{\text{lb}}=0$) to $z\approx 2.6$. }
\label{fig:train}
\end{figure*}

The vast majority of observations at redshift $z\sim 1 - 2$ have measured gas velocity dispersions of the order $\sigma_{v}\sim40$-$80\;\text{km/s}$ for the thick component of galaxy discs \citep{genzel2008,wisnioski2015}, however, those are usually clumpy disc galaxies without obvious spiral structures.  Spiral galaxies at 
similar redshifts likely have lower velocity dispersions ($\sigma_{v}\sim20$-$40\;\text{km/s}$) \citep{yuan2017,diteodoro2018}, with merger-triggered spirals representing an exception \citep{law2012}. All nearby spirals systematically show 
low gas velocity dispersion and disc scale height \citep{epinat2010}. Our simulation is consistent with the idea that long-lived density wave spiral arms reside in low-velocity dispersion thin discs \citep{yuan2017}.

\subsection{Kinematical properties of the stellar populations on the galaxy disc} \label{sec:gas}

In Fig. \ref{fig:thickness}, we show how the full width at half maximum (FWHM) of the height distribution of the stellar populations in the galaxy, $\mathcal{P}(\lvert z_{\text{th}} \rvert)$, 
varies as a function of the age (top panel) and metallicity (bottom panel) of the galaxy stellar populations 
at redshift $z=2$. 
The black triangle in the top panel of Fig. \ref{fig:thickness} corresponds to the FWHM of the height distribution of the galaxy gas particles, 
which are predicted to reside on a very thin disc at $z=2$. 

First of all, by looking at Fig. \ref{fig:thickness}, as we consider older stellar populations, they 
typically cover larger ranges of galactic altitudes. Secondly, also the stellar metallicity 
is strongly correlated with the height distribution of the stars in the galaxy, with the metal-poor stars covering much wider ranges 
of galactic altitudes than the metal-rich stars, which are typically concentrated on thinner discs. Both age and metallicity dependencies are consistent with observations in the Milky Way \citep{ting2018,mackereth2019}, 
and also with predictions of the monolithic collapse scenario \citep[e.g.,][]{larson1974} and chemodynamical simulations \citep[e.g.,][]{kobayashi2011a}.

In Fig. \ref{fig:j-sigma_figure}, stellar populations of different ages are disentangled to show how their main kinematical properties vary 
as functions of their galactocentric distance. In particular, the various panels show the radial profile of the specific angular momentum 
of stars in different age bins (top panel), 
the radial profile of the rotational velocity dispersion (middle panel), and the radial profile of the vertical velocity dispersion (bottom panel). 
Fig. \ref{fig:j-sigma_figure} shows  that the old stellar populations 
are more dominated and supported by their random motions, having -- at any galactocentric distance -- lower specific angular momenta, 
and higher rotational and vertical velocity dispersions, than the young stellar populations. It is worth noting that 
the youngest stellar populations have very similar kinematical properties 
as the gas in the galaxy, at any galactocentric distances, since they both determine galaxy structures which are rotation-supported. The exponential profiles of the vertical velocity dispersion versus radius relation  are consistent with  local spiral galaxies (e.g., \citealt{aniyan2018}).

\subsection{Properties of the stars on the spiral arms} \label{sec:arms}

In Fig. \ref{fig:SFR-maps}, we show where the star formation activity takes place in our simulated disc galaxy, from $z=3$ down to $z=2$. In the background, with black dots, we show 
the spatial distribution of the gas particles in the galaxy, viewed edge-on, and we highlight in yellow all the star-forming gas particles in the galaxy. We find that the SFR is highest in the bar at all redshifts.
The SFR is also high in clumps along the spiral arms of the galaxy.
We also find more intense star formation activity along the spiral arms as the galaxy approaches redshift $z=2$, where there is the peak in the cosmic star formation rate \citep{madau2014}.
Our findings are in agreement with the observations in the local Universe, where the star formation activity 
usually takes place in dense molecular clouds along spiral arms \citep{schinnerer2013}.
In the present-day bulges, however, we do not have any evident sign of ongoing strong star formation activity in nearby disc galaxies, even though 
bulges are usually heavily obscured by dust extinction \citep{nelson2018}.

Unlike the instantaneous snapshots of observed galaxies (\S 1), our simulated galaxy allows us to probe the origin of spiral arms.
We did this by tracing the ID numbers of star particles, following the evolution of the spiral structures in the simulated galaxy.
In Fig. \ref{fig:f_spiral}, we quantify the fraction of the stellar populations residing on the spiral arms at different redshifts, by identifying the star particles on the spiral arms at different redshifts. 
If at two different redshifts we can prove that there are very different populations of stars on the spiral arms, then this is a signature of the fact 
that spiral arms in our simulated galaxy originate from density wave perturbations propagating on the galaxy disc, and that the spiral arms are not induced by mergers or accretion events. 

The $y$-axis in Fig. \ref{fig:f_spiral} represents 
the fraction of stars on the spiral arms, in common between the time indicated by each colour (corresponding to the redshifts $z=2.0$, $2.2$, and $2.4$) and a set of 
previous epochs of the galaxy evolution. The $x$-axis of Fig. \ref{fig:f_spiral} represents the look-back time, starting from the time corresponding to the 
redshift when we compute $f_{\text{spiral}}$. In Fig. \ref{fig:f_spiral}, 
we show the evolution of $f_{\text{spiral}}$ as a function the look-back time, by considering only the stellar populations with B-band luminosity,  $ l_{\text{B},\star}$, 
satisfying the following condition:
\begin{equation} \log( l_{\text{B},\star} / l_{\text{B},\sun} ) > {3.5}. \label{eq:lb} \end{equation} 
Even if we considered either all the stellar populations on the arms (regardless their age, metallicity, B-band luminosity, and so on), 
or the stellar populations following equation \ref{eq:lb}, we find results very similar as those shown in Fig. \ref{fig:f_spiral}. 

Fig. \ref{fig:f_spiral} shows that -- as the spiral pattern rotates -- there are always different stellar populations on the spiral arms.  
In particular, the fraction of stars on the spiral arms in common between different redshifts, $f_{\text{spiral}}$, 
rapidly decreases as a function of the look-back time, being $< 15$ per cent for 
$\Delta t \approx 0.1 \,\text{Gyr}$, and as low as $5$ per cent for $\Delta t \approx 0.5 \,\text{Gyr}$, whatever be the initial redshift we consider for reference. If we fit with a decaying exponential function $f_\text{spiral}$ as a function of the look-back time in Fig. \ref{fig:f_spiral}, we predict a decay time-scale $\tau_{\text{spiral}}\approx 193\,\text{Myr}$, over which the stellar populations typically leave the spiral arm, where they were born. 

We remark on the fact that the young stellar populations on the spiral arms have initially very similar kinematics as the gas on the galaxy disc (see Fig. \ref{fig:velocityz2}, where stars with ages $<0.5\,\text{Gyr}$ have similar kinematical properties as the gas), with an average radial velocity component which is consistent with $\approx 0\,\text{km}\,\text{s}^{-1}$ (see Fig. \ref{fig:j-sigma_figure}); therefore, the evolution of $f_{\text{spiral}}$ as a function of the look-back time in Fig. \ref{fig:f_spiral} is not an artifact of the epicyclic motion of stars. 

In conclusion, we find that the spiral arms in our simulated high-redshift disc galaxy typically 
host star-forming gas particles and young stellar populations; moreover, spiral arms are like a perturbation, which propagates on the galaxy disc 
with a different angular velocity than that of the gas 
and stars on the disc. 

\section{The angular velocity of the spiral arms} \label{sec:arms}

Whether we are dealing with classical (e.g., \citealt{lin1964}) or kinematical (e.g., \citealt{dobbs2010}) spiral density waves or with manifold-driven spiral arms \citep{athanassoula2012}, the angular velocity of the spiral pattern perturbation should appear constant as a function of galactocentric distance. In other words, while the stars and gas on the disc show differential angular rotation, the spiral pattern should move 
like a solid body, with constant radial profile for the angular rotation velocity.  
In this Section, we investigate all these aspects by looking at the properties of the spiral pattern at different epochs of the galaxy evolution in our simulation. 

First of all, we place many ideal observers at different galactocentric distances, $r$, and azimuth angles, $\theta$, sitting at rest on the galaxy disc; the position of these observers is shown in Fig. \ref{fig:eobservers}. Each individual observer then registers the train of impulses as a function of time, as due to the passage of the spiral arm density perturbation. We assume that each observer defines a box region around them with size $0.1\,\text{kpc}$.

In Figure \ref{fig:train}, we show the time evolution of the train of impulses in the gas density, as registered by the observers of Fig. \ref{fig:eobservers}, covering different azimuth angles and distances on the galaxy disc. We consider all available snapshots, from redshift $z\approx1.8$ to $z\approx2.6$, with the zero point for the look-back time corresponding to redshift $z\approx1.8$. Each panel in Fig. \ref{fig:train} corresponds to different galactocentric distances, and -- within each panel -- the colour-coding represents the normalised number of particles, $\mathcal{N}(r,\theta,t)$, that each observer sees at the given look-back time, where the normalisation factor, $\mathcal{A}(r,\theta)$, corresponds to the maximum impulse in the density perturbation that each observer has ever seen in the considered redshift range.

We can use Fig. \ref{fig:train} to compute the angular velocity of the spiral pattern: 
\begin{equation}
\Omega_{\text{spiral}}(r) = \Big( \frac{ d \theta (r) }{ dt } \Big)_{\text{spiral}},
\end{equation}
which may depend -- in principle -- on the galactocentric distance and time. We note that, just by overplotting the different panels of Fig. \ref{fig:train}, we can see that the various patterns of $\theta$ versus time match with each other for different galactocentric distances, meaning -- qualitatively -- that the spiral structure perturbation rotates on the disc like a solid body; nevertheless, we would like to develop a simple analysis to quantify more precisely  $\Omega_{\text{spiral}}(r)$ on the galaxy disc.

\begin{figure}
\centering
\includegraphics[width=8.0cm]{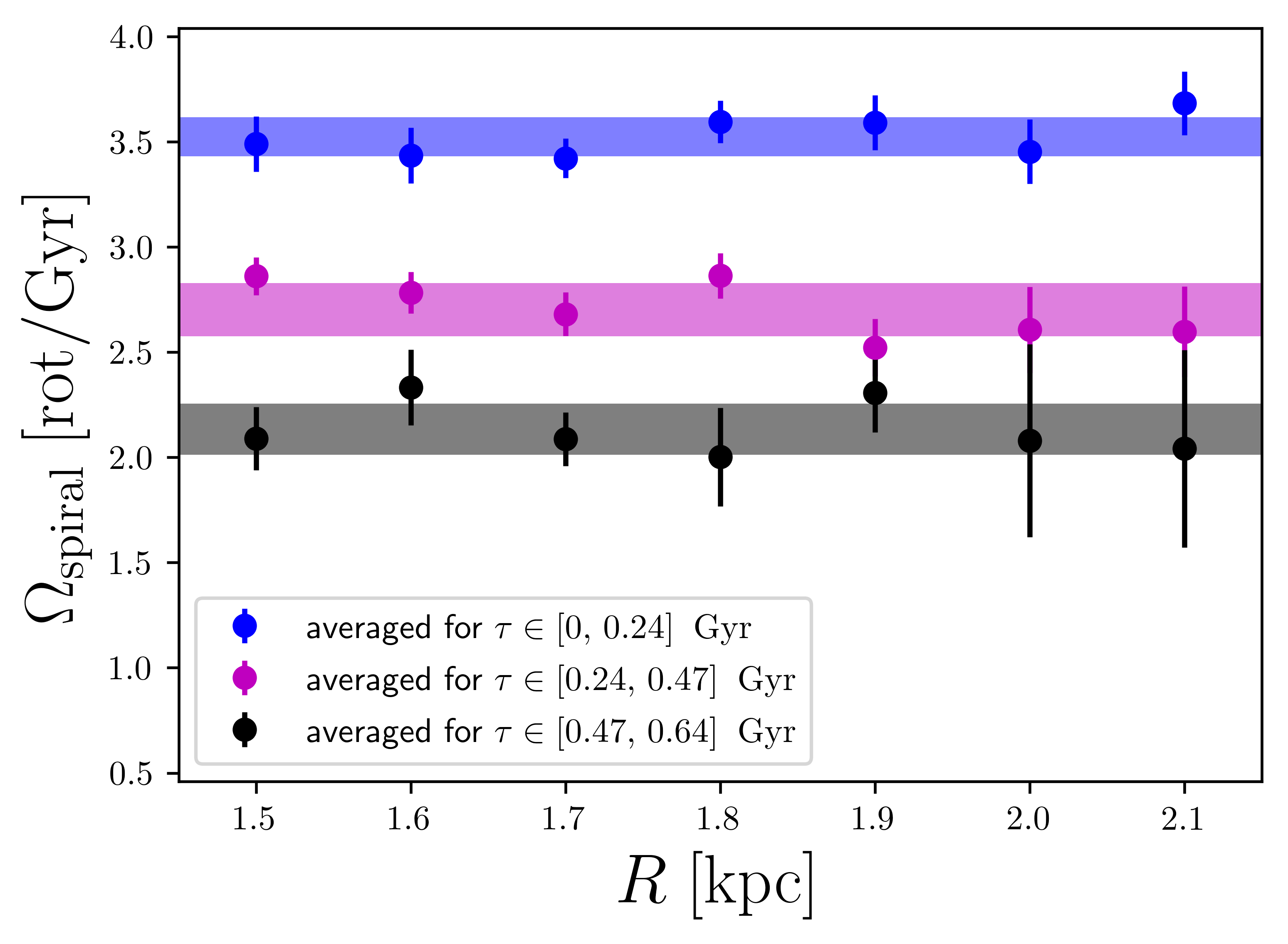} 
\caption{The time evolution of the spiral pattern angular velocity as a function of 
the galactocentric distance, as determined from 
the analysis of Fig. \ref{fig:train}, by computing the slopes of $\theta(r)$ versus time, in different ranges of time in the past, for different galactocentric distances, $r$.   }
\label{fig:angular_spiral_evolution}
\end{figure}

In order to quantify $\Omega_{\text{spiral}}(r)$, we have computed the slopes of $\theta$ versus $t$ in Fig. \ref{fig:train}, by fitting -- for different galactocentric distances -- the predicted $\theta$--time relations of the overdensities, assuming a simple linear relation. The time evolution of $\Omega_{\text{spiral}}(r)$ is then simply obtained by fitting the predicted $\theta$--time relations of the overdensities in different time intervals. The results of our analysis are shown in Fig. \ref{fig:angular_spiral_evolution}, where different colours correspond to different intervals in the look-back time. Finally, the error bars in Fig. \ref{fig:angular_spiral_evolution} correspond to the errors in the best fit parameters, with the shaded coloured areas representing the average angular velocity (with the corresponding $\pm1\sigma$ deviations), in the considered time interval. 

The main findings of our analysis are that \textit{(i)} $\Omega_{\text{spiral}}$ is almost constant on the main body of the galaxy gaseous disc;  \textit{(ii)} $\Omega_{\text{spiral}}$ increases, on average, as a function of time, which is consistent with the increase of the specific angular momentum as a function of time, seen in Fig. \ref{fig:evolution_j-sigma_gas}. 
The pattern speed of spiral arms is considered an important feature in secular evolution of disc galaxies \citep{Buta2009} and is related to the  angular momentum transport within the disc \citep{Lynden-Bell1972}.   
In our simulated galaxy, we have continuous gas accretion, which deposit kinetic energy and momentum onto the galaxy disc, enhancing its specific angular momentum as a function of time, particularly in the outer galaxy regions, which are less gravitationally bound; at the same time, the spiral pattern perturbation keeps its constant radial profile as a function of radius.   

We note that our estimated values for $\Omega_{\text{spiral}}$ rely on the assumption that we can fit the predicted $\theta$ versus $t$ relations of the overdensities with a simple linear relation; this is the main source of error in our analysis, since it is clear -- just by looking at Fig. \ref{fig:train} -- that the slope of $\theta$ versus $t$ slightly changes with time, particularly at high-redshifts, when the galaxy disc is developing; moreover, there are also many local, sudden inhomogeneities appearing on the galaxy disc. All these effects are taken into account in the error estimate of $\Omega_{\text{spiral}}$. Nevertheless, we acknowledge that a more precise analysis -- measuring the instantaneous $\Omega_{\text{spiral}}$ -- would link the evolution of $\Omega_{\text{spiral}}$ to disturbance events from external (e.g., accretion of substructures through the filaments) or internal sources (e.g., star formation activity and feedback). 

\begin{figure}
\centering
\includegraphics[width=8.0cm]{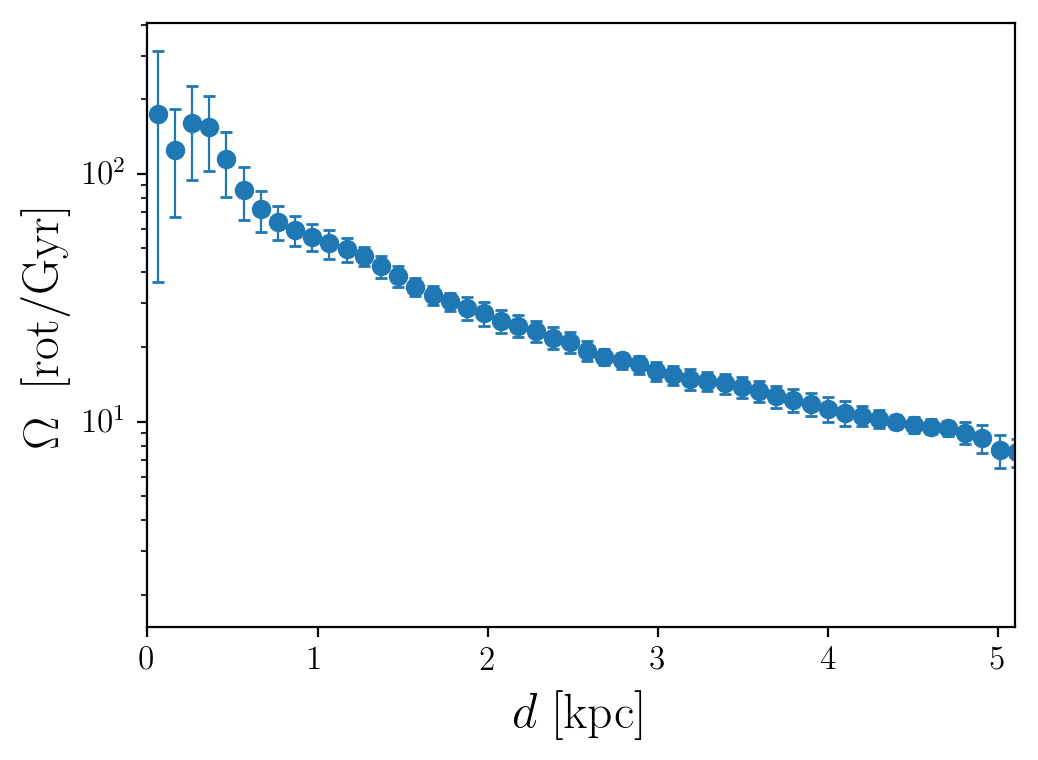} 
\caption{The radial profile of the angular velocity of the gas (blue points with errorbars) on the galaxy disc of our simulated disc galaxy at redshift $z\approx2$. }
\label{fig:angular_velocity_profile}
\end{figure}

\begin{figure*}
\centering
\includegraphics[width=16.0cm]{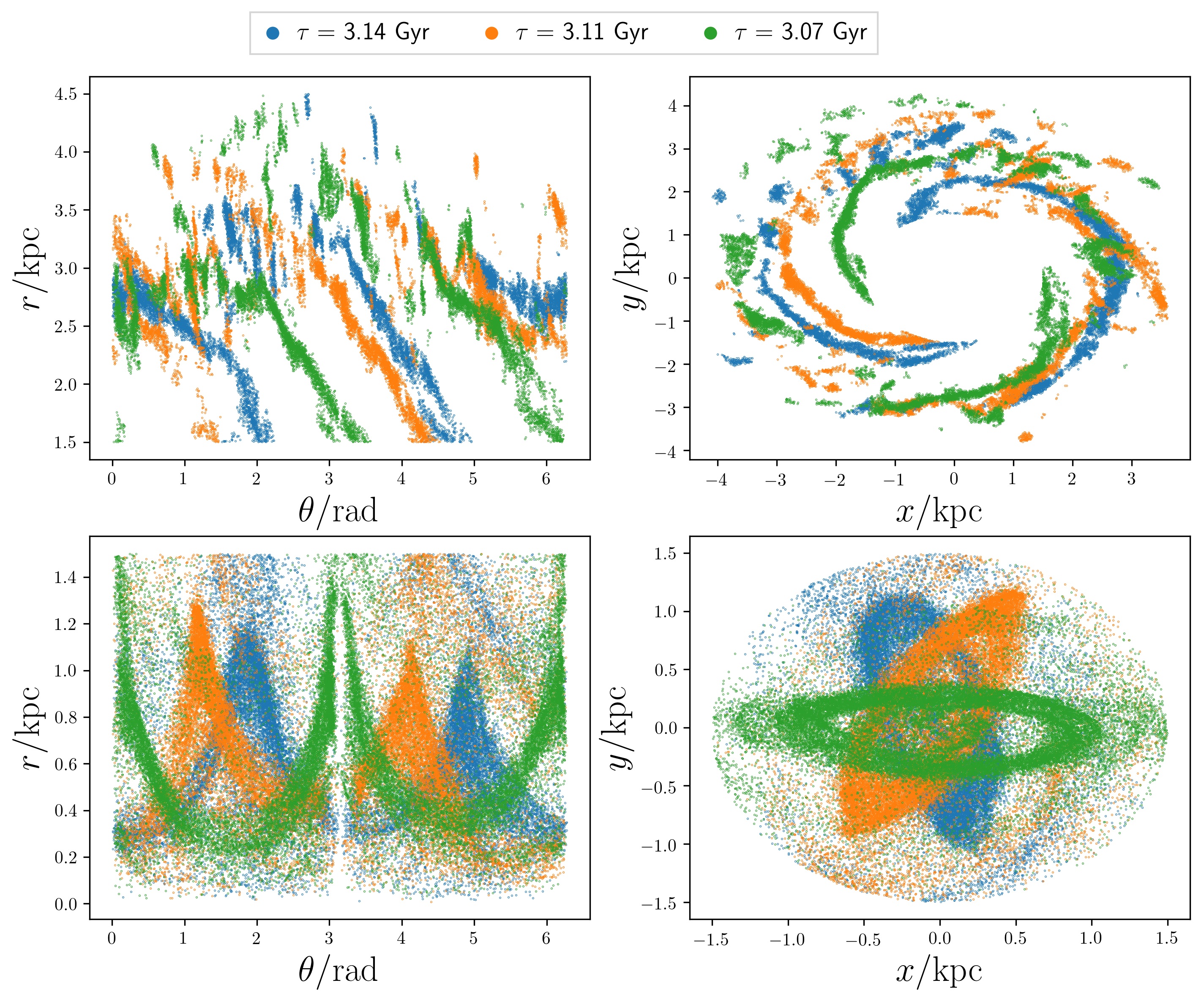} 
\caption{The first column shows the spiral arms (top panel) and the bar (bottom panel) at 
different epochs of the galaxy evolution 
in the $r$-$\theta$ diagram, where $r$ represents the galactocentric 
distance and $\theta$ the azimuthal angle. The second column shows the spiral arms (top panel) and the bar (bottom panel) 
at three epochs (blue, orange, and red points), with the galaxy being viewed face-on in the $x$-$y$ plane.  }
\label{fig:evolution_spiralpattern_bar}
\end{figure*}

The angular velocity of the spiral pattern (see Fig. \ref{fig:angular_spiral_evolution}) is much slower than the circular motion of the gas on the disc, which is shown in Fig. \ref{fig:angular_velocity_profile}. This can be better 
appreciated by computing the co-rotation radius,  $r_{\text{cr}}$, which is defined as the galactocentric distance, where the following equivalence is satisfied.

\begin{equation}
\Omega_{\star}(r_{\text{cr}}) = \frac{v_{\text{circ}}}{r_{\text{cr}}}  \equiv \Big( \frac{ d \theta }{ dt } \Big)_{\text{spiral}},
\label{eq2}
\end{equation}
where $\Omega_{\star}$ represents the angular velocity of the stellar populations on the galaxy disc. 
If we assumed a flat rotation curve which extends well beyond the physical dimensions of our simulated galaxy at $z=2$, 
then we would have a co-rotation radius $r_{\text{cr}}\approx13\,\text{kpc}$. 

We notice from Fig.~\ref{fig:evolution_fields} and Fig.~\ref{fig:evolution_j-sigma_gas} that 
 the  spiral pattern changes from less regular to fully organised   when the gas settles from a thick (with vertical dispersion $\sigma_{v} >$ 50 km/s) to a thin ($\sigma_{v} \sim 25$ km/s) disc component from $z\sim 3$ to $z\sim 2$. This co-evolution of spiral arms and thin discs is expected in density wave theories of spiral arm formation (e.g., \citealt{lin1964,bottema2003,sellwood2014}). 

Manifold theory \citep{athanassoula2012} may also be a good alternative explanation for the origin of the spiral arms, and for the constancy of their angular velocity as a function of radius, since the spiral pattern in our simulation is seen -- at all redshifts -- to co-rotate with the strong central bar (see Fig. \ref{fig:evolution_spiralpattern_bar}). In this scenario, the gas naturally accumulates in invariant manifolds, which are defined as regions around the unstable Lagrange points of the bar, co-rotating with the bar. Nevertheless, we remark on the fact that the gas and stars particles in our simulation have intrinsically higher angular velocity than the spiral pattern at the same radius (see Fig. \ref{fig:angular_velocity_profile}). A more careful analysis is needed to disentangle between the kinematic density-wave theory and the manifold theory, because both of them could give rise to a constant pattern speed as a function of radius. We leave this study to a future work, in which we will perform accurate orbital analysis of the star particles in our simulations, by also making use of other available codes (e.g., \citealt{bovy2015}).

\section{Conclusions} \label{sec:conclusions}

In this paper, we have presented the results of our zoom-in cosmological chemodynamical simulation, which unintentionally demonstrated the formation and evolution of a star-forming, barred spiral galaxy from redshift $z\sim3$ to $z\sim2$.

At redshift $z=2$, the simulated galaxy in our zoom-in simulation has a total stellar mass $M_{\star}=4.16\times10^{10}\,\text{M}_{\sun}$, total gas mass $M_{\text{gas}}=9.03\times10^{9}\,\text{M}_{\sun}$, V-band half-light radius $\sim 3$ kpc, disc scale height $\sim 0.36$ kpc, and the
average circular velocity $\langle v_{\text{circ}} = \rangle 287.0 \pm 12.3 \, \text{km/s}$, and hence it is a thin disc galaxy.
The average stellar and gas-phase metallicities are $\log(Z_{\star}/Z_{\sun})=-0.24$  and $\log(Z_{\text{gas}}/Z_{\sun})=-0.30$, respectively,   
and the metallicity evolution is driven by a large-scale primordial gas accretion (Fig. 3). 
Our study demonstrates that high-resolution cosmological hydrodynamical simulations are 
now ready to examine the formation and evolution of high-redshift spiral galaxies in the same detailed manner as of nearby spiral galaxies. Our main conclusions can be summarised as follows.

\begin{enumerate}
    \item The seed of our simulated disc galaxy is represented by the central ``bulge'', which formed at high redshift from  the assembly of many dense, gas-rich clumps. Following the size growth of the disc, the galaxy develops from redshift $z\approx 3.5$, with two trailing spiral arms being present down to redshift $z \approx 1.8$. 
    \item  By identifying star particles during the evolution of the spiral structure, we find that the spiral arms originate from density wave perturbations.
    The stellar populations newly-born in the arms leave the spiral arms over an average typical time-scale $\tau_{\text{spiral}}\approx 193\,\text{Myr}$, irrespective of redshift (Fig. 11).
    \item The pattern of the spiral arms rotates like a solid body, propagating like a perturbation on the galaxy disc, with relatively low constant angular velocity (approximately three rotations per Gyr), setting the galaxy co-rotation radius at $13.6\,\text{kpc}$ (Fig. 12).     
    \item The central bulge is constituted by an X-shaped bar, with the orbits of the particles in the simulation following a ``figure of eight'' as a function of time. The bar formed by $z \sim 3$ and the presence of the bar determines an increase (i.e. a heating) of the radial velocities of the gas particles in the galaxy centre. 
    \item The star formation activity in our simulated disc galaxy takes place in the central bulge and in several clumps on the spiral arms, at all redshifts. The number and size of the star-forming gas clumps on the arms increases as a function of redshift, reaching a maximum at $z\approx 2$, when we have the peak of the cosmic SFR. 
    \item By analysing the kinematic properties of stellar populations with different age in the galaxy, we find that stellar populations with increasing age are concentrated, on average, towards higher galactic latitudes and have also lower average metallicities. 
    The old stellar populations have lower specific angular momentum and higher velocity dispersion than the young ones, at all galactocentric distances. 
    \item The specific angular momentum of the gas on the galaxy thin disc increases as a function of redshift,
    with the angular velocity of the spiral pattern, $\Omega_{\text{spiral}}$, keeping always its radial profile constant.
    The velocity dispersion in the thin disc remains always lower, on average, than $\sim 50\,\text{km/s}$. 
    \item The dynamical structure of the spiral arms in the simulation is different than that of the bulk of the stars and gas on the galaxy disc; in particular, $\Omega_{\text{spiral}}$ increases as function of time (like the specific angular momentum of the gas in the disc), maintaining a profile which is constant as a function of radius. This suggests that the spiral pattern is a fundamental process with which angular momentum is transported within our simulated disc galaxy \citep{Lynden-Bell1972}. Nevertheless, with our current analysis, we cannot robustly disentangle between kinematic density waves and manifold theory for the origin of the spiral arms, because both theories can give rise to a constant radial profile of the angular velocity of the spiral pattern on the galaxy disc; this will be the subject of our future work 
\end{enumerate}


\section*{Acknowledgments}
We thank an anonymous referee for many insightful and thought-provoking comments, which improved the quality and clarity of our work. We thank Benjamin Davis, Andrew Bunker, Roberto Maiolino, and Francesco Belfiore for many interesting and stimulating discussions. 
FV and CK acknowledge funding from the United Kingdom 
Science and Technology Facility Council through grant ST/M000958/1 and ST/R000905/1. 
TY acknowledges a Fellowship support from the Australian Research Council Centre of Excellence for All Sky Astrophysics in 3 Dimensions (ASTRO 3D), through project number CE170100013.
FV and TY also thank all the participants of the 
workshop ``Gas Fuelling of Galaxy Structures
across Cosmic Time'', held between 5th and 9th November 2018 in 
Barrossa Valley, South Australia, for the many discussions. 
FV acknowledges support from the European Research Council Consolidator Grant funding scheme (project ASTEROCHRONOMETRY, 
G.A. n. 772293). 
FV thanks ASTRO 3D
for kindly supporting his visit at the Swinburne University of Technology in 2018 November.
This work used the DiRAC Data Centric system at Durham University, operated by the Institute for Computational Cosmology on behalf of the STFC DiRAC HPC Facility (www.dirac.ac.uk). This equipment was funded by a BIS National E-infrastructure capital grant ST/K00042X/1, STFC capital grant ST/K00087X/1, DiRAC Operations grant ST/K003267/1 and Durham University. DiRAC is part of the National E-Infrastructure.
This research has also made use of the University of Hertfordshire's high-performance computing facility.  
We finally thank Volker Springel for providing the code 
\textsc{Gadget-3}.






\end{document}